# Four-point renormalized coupling constant in $O(N)$ models.

Massimo Campostrini, Andrea Pelissetto, Paolo Rossi, and
Ettore Vicari
*Dipartimento di Fisica dell'Università and I.N.F.N., I-56126 Pisa, Italy*

The renormalized zero-momentum four-point coupling $g_r$ of $O(N)$-invariant scalar field theories in $d$ dimensions is studied by applying the $1/N$ expansion and strong coupling analysis.

The $O(1/N)$ correction to the $\beta$-function and to the fixed point value $g_r^*$ are explictly computed. Strong coupling series for lattice non-linear $\sigma$ models are analyzed near criticality in $d = 2$ and $d = 3$ for several values of $N$ and the corresponding values of $g_r^*$ are extracted.

Large-$N$ and strong coupling results are compared with each other, finding a good general agreement. For small $N$ the strong coupling analysis in 2-d gives the best determination of $g_r^*$ to date (or comparable for $N = 2, 3$ with the available Monte Carlo estimates), and in 3-d it is consistent with available $\phi^4$ field theory results.

## I. INTRODUCTION.

The study of the fixed-point behavior of quantum and statistical field theories is one of the central problems to be faced both from a purely theoretical point of view and for the purpose of investigating such phenomenologically relevant issues as the existence and quantitative estimate of triviality bounds.

For understandable reasons most theoretical effort has been till now directed towards the analysis of a few selected models, including $O(0)$, $O(1)$, $O(2)$ and $O(3)$ in three dimensions and $O(4)$ ("Higgs") models in four dimensions. In our view, it is very useful to extend the analysis to the case of a generic symmetry group $O(N)$ and to models living in an arbitrary number of space dimensions $d$, not only for the sake of generality or in view of new possible physical applications, but also because this analysis may offer the possibility of testing and cross-checking the several different methods that have been applied to the problem at hand, thus putting on firmer grounds results that often rely only on a single approach and whose generality cannot therefore be fully understood.

In making these statements, we have especially in mind the possibility of systematically extending the application of two very well known techniques: strong coupling and $1/N$ expansion. Despite their all too evident advantages, i.e. finite convergence radius for the strong coupling expansion and non-perturbative interpretation of results for the $1/N$ expansion, these techniques suffer from some drawbacks, i.e. lack of control on the accuracy of the resummation techniques in the strong coupling case, poor information on the convergence properties and technical difficulty in the extending the series for the $1/N$ expansion, which have often discouraged people from pursuing these approachs.

Neverthless we think, and we hope to show convincingly in the present paper, that a renewed effort in these directions may prove very fruitful, especially when two classes of results are systematically compared, since agreement of $1/N$ expansion and strong coupling results for a given $N$ can be taken as evidence of convergence of the $1/N$ expansion down to that value of $N$, while the non-perturbative character of the $1/N$ expansion ensures the generality of the qualitative results (like triviality) within all the convergence domain.

In Sec. II we introduce our notation for the general $O(N)$-invariant scalar field theory with quartic interaction in the continuum formulation and define the quantity which we are going to study, i.e. the renormalized zero-momentum four-point coupling $g_r$, whose behavior in the scaling region and fixed-point value for arbitrary $N$ and $d$ is the object of our investigations. This quantity is related to the so-called Binder cumulant. We compute the next-to-leading $1/N$ correction to the renormalized coupling as a function of the bare coupling and the renormalized mass by reducing it to a set of Feynman integrals that can be evaluated in the continuum $1/N$ expanded model without any regularization for all $d < 4$. We show how to compute the $1/N$ correction to the $\beta$-function $\beta(g_r) = m_r \frac{dg_r}{dm_r}$ and the fixed point value $g_r^*$ such that $\beta(g_r^*) = 0$.

In Sec. III we give an exact evaluation of $g_r$ in one dimension by solving exactly the one-dimensional non-linear $\sigma$ models for arbitrary $N$, and draw from this example some general indication about the possible dependence of $g_r^*$ on the parameter $N$.

In Sec. IV we review the available results on non-linear $\sigma$ models in two and three dimensions on the lattice and present our explicit computations of $1/N$ effects for the above models at criticality. We briefly comment on the case $d = 4$ and $O(1/N)$ logarithmic deviations from scaling.

In Sec. V we analyze and discuss the strong coupling series for the renormalized coupling of the non-linear $\sigma$ models in two and three dimensions on the lattice and for arbitrary $N$, which we extracted from the results of Lüscher and Weisz [1] as elaborated by Butera et al. [2]. Strong coupling results are compared to all available calculations presented in the literature ($\phi^4$ field theory at fixed dimensions and Monte Carlo simulations) and to our $1/N$ results, finding a good general agreement.

Finally in Sec. VI we draw some conclusions.

An explicit representation of the $1/N$ correction to the $\beta$-function is exhibited and discussed in the Appendix.

## II. THE RENORMALIZED COUPLING AND ITS $1/N$ EXPANSION.

According to the previous discussion it is interesting to form a renormalization-group invariant dimensionless combination of vacuum expectation values playing the rôle of a renormalized four-point coupling and to study its behavior in the proximity of a critical point. In particular we are interested in $O(N)$-invariant scalar field theories in arbitrary dimensions $d \leq 4$ and we wish to apply $1/N$ expansion techniques to the above-mentioned problem.

From the point of view of the $1/N$ expansion the standard notation is somewhat inconvenient: we shall therefore define our own conventions, trying to establish correspondence with the literature as far as possible, and especially trying to make all relationships with Refs. [1,3–6] as transparent as we could.

The usual $O(N)$-invariant Euclidean continuum Lagrangian takes the form

$$\mathcal{L} \;=\; \frac{1}{2}\partial_\mu\vec{\phi}\partial_\mu\vec{\phi} + \frac{1}{2}\mu_0^2\vec{\phi}^2 + \frac{g_0}{4!}(\vec{\phi}^2)^2 \;. \tag{1}$$

It is however convenient to redefine the quartic coupling (both bare and renormalized) according to the definition

$$\widehat{g} \;=\; \frac{Ng}{3} \;. \tag{2}$$

We shall also define

$$\beta \equiv \frac{-2\mu_0^2}{\widehat{g}_0} \;, \qquad \gamma \equiv \frac{1}{\widehat{g}_0} \;, \qquad \vec{s} \equiv \frac{\vec{\phi}}{\sqrt{N\beta}} \;, \tag{3}$$

and introduce an auxiliary field $\alpha$.

The resulting effective Lagrangian is

$$\mathcal{L} \;=\; \frac{N}{2}\left[\beta\partial_\mu\vec{s}\partial_\mu\vec{s} + i\beta\alpha(\vec{s}^{\,2}-1) + \gamma\alpha^2\right] \;, \tag{4}$$

and after performing a Gaussian integration over the field $\vec{s}$ we obtain

$$\mathcal{L} \;=\; \frac{N}{2}\left[\mathrm{Tr}\ln\beta\left(-\partial_\mu\partial_\mu + i\alpha\right) \;-\; i\beta\alpha \;+\; \gamma\alpha^2\right] \;, \tag{5}$$

which reduces to the usual effective large-$N$ action for the non-linear $\sigma$ model in the limit $\gamma \to 0$.

Correspondence with Refs. [1,3–6] is established by the relationships

$$\beta = \frac{2\kappa}{N} \;, \qquad \gamma = \frac{1}{N}\frac{\kappa^2}{2\lambda} \;. \tag{6}$$

Renormalization is performed according to the following prescriptions for the two and four-point correlation functions of the field $\vec{\phi}$:

$$\Gamma^{(2)}(p,-p)_{\alpha\beta} = \Gamma^{(2)}(p^2)\delta_{\alpha\beta} \;, \qquad \Gamma^{(2)}(p^2) = \; Z_r^{-1}\left[m_r^2 + p^2 + O(p^4)\right] \;, \tag{7}$$

and

$$\Gamma^{(4)}(0,0,0,0)_{\alpha\beta\gamma\delta} = \Gamma^{(4)}(0)\left(\delta_{\alpha\beta}\delta_{\gamma\delta} + \delta_{\alpha\gamma}\delta_{\beta\delta} + \delta_{\alpha\delta}\delta_{\beta\gamma}\right) \;, \qquad \Gamma^{(4)}(0) = -Z_r^{-2}\frac{\widehat{g}_r}{N}(m_r^2)^{2-\frac{d}{2}} \;. \tag{8}$$

Let's now notice that

$$-N\frac{\Gamma^{(4)}(0,0,0,0)_{\alpha\alpha\gamma\gamma}}{\left[\Gamma^{(2)}(0,0)_{\alpha\alpha}\right]^2}(m_r^2)^{2-\frac{d}{2}} \;=\; \left(1+\frac{2}{N}\right)\widehat{g}_r \;. \tag{9}$$

Equation (9) will be our working definition of the renormalized four-point coupling.

In order to compute the leading and next-to-leading contributions to $\widehat{g}_r$ in the continuum $1/N$ expansion, we shall need an evaluation of the corresponding contribution to the two-point function and to the zero-momentum four-point function.

The evaluation of the Feynman rules shown in Fig. 1 is essentially straightforward. We only mention that the bare propagator of the $\vec{\phi}$ field is expressed in terms of a "bare" large-$N$ mass parameter $m_0^2$ introduced by the gap equation

$$\int \frac{d^dp}{(2\pi)^d}\frac{1}{p^2+m_0^2} \;=\; \beta + 2\gamma m_0^2 \;, \tag{10}$$

while the propagator of the Lagrange multiplier field $\alpha$ is defined to be $\frac{1}{N}\Delta(k,m_0^2)$, where in turn

$$\begin{aligned}\Delta^{-1}(k,m_0^2) &= \frac{1}{2}\int\frac{d^dp}{(2\pi)^d}\frac{1}{p^2+m_0^2}\,\frac{1}{(p+k)^2+m_0^2} \;+\; \gamma \\ &= \frac{1}{2}\frac{\Gamma\left(2-\frac{d}{2}\right)}{(4\pi)^{d/2}}\left(\frac{k^2}{4}+m_0^2\right)^{\frac{d}{2}-2} F\left[2-\frac{d}{2},\frac{1}{2},\frac{3}{2},\left(1+\frac{4m_0^2}{k^2}\right)^{-1}\right] \;+\; \gamma \;.\end{aligned} \tag{11}$$

The relevant higher-order lagrangian effective vertices are obtained by taking derivatives of $\Delta^{-1}(k)$ with respect to $m_0^2$, according to the correspondence table

$$
\begin{aligned}
V^{(3)}(0,k,k) &= -\frac{\partial}{\partial m_0^2}\Delta^{-1}(k)\,, \\
2V^{(4)}(0,0,k,k)+V^{(4)}(0,k,0,k) &= -\frac{\partial}{\partial m_0^2}V^{(3)}(0,k,k) = \frac{\partial^2}{(\partial m_0^2)^2}\Delta^{-1}(k)\,,
\end{aligned}
\tag{12}
$$

where mass dependence is suppressed in the arguments. The derivatives appearing in Eq. (12) may be evaluated by a generalization of the so-called "cutting rule" of Ref. [7], whose $d$-dimensional form is

$$
\frac{\partial}{\partial m_0^2}\Delta^{-1}(k) = -\frac{2}{k^2+4m_0^2}\left[(3-d)\Delta^{-1}(k)+\Delta^{-1}(0)+\gamma(d-4)\right]\,. \tag{13}
$$

In writing Eqs. (10) and (11) some ultraviolet regularization, when needed, is assumed. Actually our final results will turn out to be independent of the regularization as expected on physical grounds.

Equation (9) shows that in order to compute $\widehat{g}_r$ to any definite order in the $1/N$ expansion we must be able to compute the quantities $\Gamma^{(2)}(p)$, $\Gamma^{(4)}(0)$ and $m_r^2$ with the same precision. Leading order calculations are straightforward. Next-to-leading contributions may be formally represented in terms of a few fundamental integrals, which are graphically represented in Fig. 2 and listed below:

$$
\Sigma_1^{(a)}(p^2,m_0^2) = \int \frac{d^dk}{(2\pi)^d}\frac{\Delta(k,m_0^2)}{(p+k)^2+m_0^2}\,, \tag{14}
$$

$$
\Sigma_1^{(b)}(m_0^2) = -\frac{1}{2}\Delta(0,m_0^2)\int\frac{d^dk}{(2\pi)^d}V^{(3)}(0,k,k)\Delta(k,m_0^2)\,, \tag{15}
$$

$$
B_1(m_0^2) = \int\frac{d^dk}{(2\pi)^d}\frac{\Delta(k,m_0^2)^2}{(k^2+m_0^2)^2}\,. \tag{16}
$$

It is very easy to show that the two-point function and the renormalized mass are respectively

$$
\Gamma^{(2)}(p^2) = p^2+m_0^2+\frac{1}{N}\left[\Sigma_1^{(a)}(p^2,m_0^2)+\Sigma_1^{(b)}(m_0^2)\right]+O\left(\frac{1}{N^2}\right) \tag{17}
$$

and

$$
m_r^2 = m_0^2+\frac{1}{N}\left[\Sigma_1^{(a)}(0,m_0^2)+\Sigma_1^{(b)}(m_0^2)-m_0^2\frac{\partial\Sigma_1^{(a)}}{\partial p^2}\left(0,m_0^2\right)\right]+O\left(\frac{1}{N^2}\right)\,. \tag{18}
$$

Explicit use of Eqs. (12), leading to the graphical identities drawn in Fig. 3 allows to obtain the representation

$$-N\Gamma^{(4)}(0) \;=\; \Delta(0,m_0^2)\left[1+\frac{2}{N}\frac{\partial}{\partial m_0^2}\Sigma_1^{(a)}(0,m_0^2)+\frac{1}{N}\frac{\partial}{\partial m_0^2}\Sigma_1^{(b)}(m_0^2)\right]-\frac{2}{N}B_1(m_0^2)+O\left(\frac{1}{N^2}\right)\;. \tag{19}$$

It is now important to notice that, in order to obtain a finite result, $\widehat{g}_r$ must be expressed in terms of the renormalized mass $m_r^2$. This is achieved by inverting Eq. (18), which leads to

$$m_0^2 \;=\; m_r^2-\frac{1}{N}\left[\Sigma_1^{(a)}(0,m_r^2)+\Sigma_1^{(b)}(m_r^2)-m_r^2\frac{\partial\Sigma_1^{(a)}}{\partial p^2}\left(0,m_r^2\right)\right]+O\left(\frac{1}{N^2}\right)\;. \tag{20}$$

and as a consequence

$$\Gamma^{(2)}(p^2) \;=\; p^2+m_r^2+\frac{1}{N}\left[\Sigma_1^{(a)}(p^2,m_r^2)-\Sigma_1^{(a)}(0,m_r^2)+m_r^2\frac{\partial\Sigma_1^{(a)}}{\partial p^2}\left(0,m_r^2\right)\right]+O\left(\frac{1}{N^2}\right)\;. \tag{21}$$

and

$$\Delta(0,m_0^2) \;=\; \Delta(0,m_r^2)+\frac{1}{N}\Delta(0,m_r^2)\left[1-\gamma\Delta(0,m_r^2)\right]\left(\frac{d}{2}-2\right)\times \left[\frac{\Sigma_1^{(a)}(0,m_r^2)}{m_r^2}+\frac{\Sigma_1^{(b)}(m_r^2)}{m_r^2}-\frac{\partial\Sigma_1^{(a)}}{\partial p^2}\left(0,m_r^2\right)\right]\;. \tag{22}$$

Collecting all the above results and substituting into Eq. (9) we obtain the following representation of the renormalized coupling:

$$\widehat{g}_r \;=\; (m_r^2)^{\frac{d}{2}-2}\Delta(0)\Bigg\{1+\frac{1}{N}\left[1-\gamma\Delta(0)\right]\left(\frac{d}{2}-2\right)\left[\frac{\Sigma_1^{(a)}(0)}{m_r^2}+\frac{\Sigma_1^{(b)}}{m_r^2}-\frac{\partial\Sigma_1^{(a)}}{\partial p^2}(0)\right] +\frac{1}{N}\left[2\frac{\partial\Sigma_1^{(a)}(0)}{\partial m_r^2}+\frac{\partial\Sigma_1^{(b)}}{\partial m_r^2}-2\frac{\partial\Sigma_1^{(a)}}{\partial p^2}(0)-2\Delta^{-1}(0)B_1\right]\Bigg\}, \tag{23}$$

where all quantities are now computed with $m_0^2$ replaced by $m_r^2$ and mass dependence is suppressed in the arguments. Substituting Eqs. (14), (15) and (16) and making explicit use of Eq. (13), one obtains the following representation:

$$\begin{aligned}
\widehat{g}_r \;=\; & (m_r^2)^{\frac{d}{2}-2}\Delta(0)\Bigg\{1+\frac{1}{N}\left[1-\gamma\Delta(0)\right](3-d)2^{d-1} \\
& +\frac{1}{N}\int\frac{d^dk}{2\pi)^d}\Delta(k)\left[\frac{2m_r^2}{(k^2+m_r^2)^3}+\frac{3m_r^2}{(k^2+m_r^2)(k^2+4m_r^2)}\left(\frac{\frac{d}{2}-4}{k^2+m_r^2}-\frac{2(d-1)}{k^2+4m_r^2}\right)\right] \\
& -\frac{1}{N}(d-4)\gamma\Delta(0)\int\frac{d^dk}{2\pi)^d}\Delta(k)\frac{1}{d}\left[\frac{2m_r^2}{(k^2+m_r^2)^3}+\frac{3m_r^2}{(k^2+m_r^2)(k^2+4m_r^2)}\left(\frac{\frac{d}{2}-2}{k^2+m_r^2}+\frac{2(d-1)}{k^2+4m_r^2}\right)\right] \\
& +\frac{1}{N}(d-4)\gamma\Delta(0)\int\frac{d^dk}{2\pi)^d}\Delta(k)\frac{2(d-1)^2}{d(k^2+4m_r^2)^2} \\
& -\frac{2}{N}\Delta^{-1}(0)\int\frac{d^dk}{2\pi)^d}\frac{\Delta(k)^2}{(k^2+4m_r^2)^2}\left[\frac{3m_r^2}{(k^2+m_r^2)}-(d-4)\gamma\Delta(0)\right]^2\Bigg\}+O\left(\frac{1}{N^2}\right)\;.
\end{aligned} \tag{24}$$

Further computational simplification is achieved by making use of the following straightforward consequence of Eq. (13):

$$\Delta^{-1}(0)\Delta(k)^2 = \frac{4m_r^2+(4-d)k^2}{4m_r^2+(4-d)\gamma\Delta(0)k^2}\Delta(k) - \frac{2k^2(k^2+4m_r^2)}{4m_r^2+(4-d)\gamma\Delta(0)k^2}\frac{\partial\Delta(k)}{\partial k^2}, \tag{25}$$

which may be applied to Eq. (24) in order to get rid of the $\Delta(k)^2$ dependence in the integrand, while a partial integration may eliminate the dependence on $\frac{\partial\Delta(k)}{\partial k^2}$. It is easy to recognize that whenever $d < 4$ and $\gamma \geq 0$ all integrations are finite. The final result can be formally expressed by the relationship

$$\widehat{g}_r = \widehat{g}_r^{(0)}(x) + \frac{1}{N}\widehat{g}_r^{(1)}(x) + O\left(\frac{1}{N^2}\right), \tag{26}$$

where all dependence on the renormalized mass and the bare coupling can only come through the dimensionless combination $x \equiv m_r^{4-d}/\widehat{g}_0$. Specifically one obtains

$$\widehat{g}_r^{(0)} = \left(\frac{1}{g_*}+x\right)^{-1}, \tag{27}$$

where

$$\frac{1}{g_*} = \frac{m_r^{4-d}}{2}\int\frac{d^dp}{(2\pi)^d}\frac{1}{(p^2+m_r^2)^2} = \frac{1}{2}\frac{\Gamma\left(2-\frac{d}{2}\right)}{(4\pi)^{d/2}} \tag{28}$$

is the inverse of the large-$N$ fixed point value of the renormalized coupling.

Equation (26) is the obvious starting point for the construction of the $1/N$ expanded $\beta$-function of the model, via the relationship

$$\beta(\widehat{g}_r) = m_r\frac{d\widehat{g}_r}{dm_r} = (4-d)x\frac{d\widehat{g}_r}{dx}|_{x=x(\widehat{g}_r)} = \beta^{(0)}(\widehat{g}_r) + \frac{1}{N}\beta^{(1)}(\widehat{g}_r) + O\left(\frac{1}{N^2}\right), \tag{29}$$

where $x(\widehat{g}_r)$ is obtained by inverting the equation $\widehat{g}_r = \widehat{g}_r(x)$, and it admits in turn a $1/N$ expansion in the form

$$x(\widehat{g}_r) = x^{(0)}(\widehat{g}_r) + \frac{1}{N}x^{(1)}(\widehat{g}_r) + O\left(\frac{1}{N^2}\right). \tag{30}$$

It is easy to recognize that Eq. (27) implies

$$x^{(0)}(\widehat{g}_r) = \frac{1}{\widehat{g}_r} - \frac{1}{g_*}, \tag{31}$$

and therefore the large-$N$ limit of the $\beta$-function reduces to

$$\beta^{(0)}(\widehat{g}_r) = (d-4)\widehat{g}_r\left(1-\frac{\widehat{g}_r}{g_*}\right). \tag{32}$$

This is the standard (large-$N$) one-loop result provided that we identify

$$\frac{4-d}{g_*} = \frac{\Gamma\left(3-\frac{d}{2}\right)}{(4\pi)^{d/2}} = \beta_0 \,. \tag{33}$$

A simple consequence of Eqs. (26), (27), (30) and (31) is the relationship

$$x^{(1)}(\widehat{g}_r) = \frac{1}{\widehat{g}_r^2}\widehat{g}_r^{(1)}(x^{(0)}(\widehat{g}_r)) \,. \tag{34}$$

We may now consider the expansion of Eq. (29) in powers of $1/N$ and notice that the derivative of $\widehat{g}_r^{(1)}(x)$ with respect to $x^{(0)}$ may be exchanged with a derivative with respect to $\widehat{g}_r$. As a consequence after some manipulations we can prove the relationship

$$\beta^{(1)}(\widehat{g}_r) = \left[\beta^{(0)}(\widehat{g}_r)\right]^2 \frac{\partial}{\partial \widehat{g}_r}\left[\frac{\widehat{g}_r^{(1)}\left(x^{(0)}(\widehat{g}_r)\right)}{\beta^{(0)}(\widehat{g}_r)}\right] , \tag{35}$$

where notable simplifications occur when evaluating $\widehat{g}_r^{(1)}$ directly as a function of $\widehat{g}_r^{-1} - g_*^{-1}$. In particular

$$m_r^{d-4}\Delta(k) \longrightarrow \frac{\widehat{g}_r}{1+\widehat{g}_r\Pi(k/m_r)} , \tag{36}$$

where

$$\Pi(k/m_r) = \frac{1}{2}\int \frac{d^dp}{(2\pi)^d}\frac{m_r^{4-d}}{p^2+m_r^2}\left[\frac{1}{(p+k)^2+m_r^2} - \frac{1}{p^2+m_r^2}\right] \tag{37}$$

is a regular dimensionless function with the property $\Pi(0) = 0$ and a finite $d \to 4$ limit. As long as $d < 4$ one may show that $\beta^{(1)}(\widehat{g}_r)$ is well-defined and finite for all $0 \le \widehat{g}_r \le g_*$. We obtained an explicit integral representation of $\beta^{(1)}(\widehat{g}_r)$ for arbitrary $d$, and showed that the series expansion of such a representation in the powers of $\widehat{g}_r$ may be obtained also in the $d \to 4$ limit and reproduces all known results as long as comparison is allowed. The representation of $\beta^{(1)}(\widehat{g}_r)$ and a short discussion of its features are presented in the Appendix. The non-perturbative properties of $\beta^{(1)}(\widehat{g}_r)$ when $d \to 4$ will be analyzed and discussed in a separate publication.

For what concerns the very important issue of the fixed point of $\widehat{g}_r$, we must notice that the $\beta$-function vanishes when $x \to 0$, i.e. when

$$\widehat{g}_r^* = g_* + \frac{1}{N}\widehat{g}_r^{(1)}(0) + O\left(\frac{1}{N^2}\right) \,. \tag{38}$$

Equation (24), supplemented with Eq. (25), lends itself to an easy evaluation in the limit $x \to 0$, corresponding to the limit $\gamma \to 0$. The final result is

$$\widehat{g}_r^{(1)}(0) = g_*\left[(3-d)2^{d-1} - \int\frac{d^dk}{(2\pi)^d}\frac{\Delta_0(k)}{(k^2+m_r^2)^2}\left(\frac{4m_r^2}{k^2+m_r^2} + 9\left(\frac{d}{2}-1\right)\frac{m_r^2}{k^2+4m_r^2}\right)\right] , \tag{39}$$

where

$$\Delta_0(k) \;=\; \lim_{x\to 0} \Delta(k)\;. \tag{40}$$

Notice that the fixed-point value of the renormalized coupling may be obtained directly by computing the $\gamma \to 0$ limit of the coupling $\widehat{g}_r$ in the scaling region. However this is nothing but the value taken by $\widehat{g}_r$ in the corresponding continuum limit field theory, that is the usual non-linear $\sigma$ model in $d$-dimensions. In turn this is the limit of the lattice non-linear $\sigma$ model when $\beta \to \beta_c$, the value of the coupling such that the renormalized mass (i.e. inverse correlation length) is equal to zero in the lattice $\gamma \to 0$ limit.

## III. NON-LINEAR $\sigma$ MODELS IN ONE DIMENSION.

Before discussing the general $d$-dimensional case, let's illustrate some features of the problem by solving the simple but not trivial one dimensional case. One-dimensional non-linear $\sigma$ models are a completely integrable system, both on the lattice and in the continuum [8–10]. Indeed in any lattice formulation with nearest-neighbor interactions the two and four-point correlation functions are easily expressed in terms of two quantities that in turn are related to vacuum expectation values of the model defined on a single link.

Without belaboring on the rather trivial manipulations needed to derive these results [8,9], we simply quote that, in any $O(N)$-invariant $\sigma$ model theory satisfying the constraint $\vec{s}\cdot\vec{s} = 1$, one may write

$$\begin{aligned}
\langle\, s^a_m s^b_n \,\rangle \;&=\; B_{11}^{n-m}\frac{1}{N}\delta^{ab}\;,\\
\langle\, \left(s^a_m s^b_m - \frac{1}{N}\delta^{ab}\right)\left(s^c_n s^d_n - \frac{1}{N}\delta^{cd}\right)\,\rangle \;&=\; B_{22}^{n-m}\left[\frac{1}{N(N+2)}\delta^{abcd} - \frac{1}{N^2}\delta^{ab}\delta^{cd}\right]
\end{aligned} \tag{41}$$

for $n \geq m$, and

$$\langle\, s^a_m s^b_n s^c_p s^d_q \,\rangle_c \;=\; B_{11}^{n-m+q-p}\left[\delta^{abcd}\left(\frac{B_{22}^{p-n}}{N(N+2)} - \frac{B_{11}^{2(p-n)}}{N^2}\right) + \delta^{ab}\delta^{cd}\frac{1}{N^2}\left(B_{11}^{2(p-n)} - B_{22}^{p-n}\right)\right] \tag{42}$$

for $q \geq p \geq n \geq m$, where $m, n, p, q$ are integer numbers labelling lattice sites,

$$\delta^{abcd} \;\equiv\; \delta^{ab}\delta^{cd} + \delta^{ac}\delta^{bd} + \delta^{ad}\delta^{bc}\;, \tag{43}$$

and

$$\begin{aligned}
B_{11} &= \langle\, \vec{s}_1\cdot\vec{s}_0 \,\rangle\;,\\
B_{22} &= \frac{N\langle\, (\vec{s}_1\cdot\vec{s}_0)^2 \,\rangle - 1}{N-1}\;,
\end{aligned} \tag{44}$$

where expectation values are taken in the single-link model: $B_{11}$ and $B_{22}$ are the character coefficients in the (pseudo) character expansion of the model [8,9], or, equivalently, the coefficient of the expansion of the theory in hyperspherical harmonics. The results corresponding to different orderings of the lattice points are obtained by trivial permutations.

Zero-momentum lattice Fourier transforms can be computed as functions of $B_{11}$ and $B_{22}$ by performing trivial summations of geometric series. One may then easily recognize that

$$\Gamma^{(2)}(0)_{aa} = \frac{1+B_{11}}{1-B_{11}} \tag{45}$$

and

$$m_r^2 = \frac{(1-B_{11})^2}{B_{11}}, \tag{46}$$

while after some purely algebraic effort one obtains

$$N\Gamma^{(4)}(0)_{aabb} = 4(N-1)\frac{B_{22}}{1-B_{22}}\left(\frac{1+B_{11}}{1-B_{11}}\right)^2 - 4NB_{11}^2\frac{1+B_{11}}{(1-B_{11})^3} - \frac{16B_{11}}{(1-B_{11})^3} - 2\,. \tag{47}$$

As a consequence by applying Eq. (9) one obtains

$$\left(1+\frac{2}{N}\right)\widehat{g}_r = \frac{1}{\sqrt{B_{11}}}\left[\frac{2(1-B_{11})^3}{(1+B_{11})^2} + \frac{4NB_{11}^2}{1+B_{11}} + \frac{16B_{11}}{(1+B_{11})^2} - 4(N-1)\frac{1-B_{11}}{1-B_{22}}B_{22}\right]\,. \tag{48}$$

We now want to take the critical limit, which, as shown in Ref. [10], depends in a complicated way on the specific hamiltonian, as in one dimension there are infinitely many universality classes. We will restrict ourselves to those theories for which $m_r \to 0$. Thus Eq. (46) implies $B_{11} \to 1$. When $N \geq 1$ we have also [10] $B_{22} \to 1$, so that

$$\left(1+\frac{2}{N}\right)\widehat{g}_r^* = 2(N+2) - 4(N-1)\lim_{B_{11}\to 1}\frac{1-B_{11}}{1-B_{22}}\,. \tag{49}$$

The quantity $\frac{1-B_{11}}{1-B_{22}}$ characterizes the universality class being simply the ratio of the mass-gap in the spin-one and spin-two channels. Within the universality class corresponding to the standard continuum limit, we have

$$\lim_{B_{11}\to 1}\frac{1-B_{11}}{1-B_{22}} = \frac{N-1}{2N} \tag{50}$$

and as a consequence

$$\left(1+\frac{2}{N}\right)\widehat{g}_r^* = 8\left(1-\frac{1}{4N}\right) \tag{51}$$

for $N \geq 1$. This solution agrees perfectly with the prediction resulting from Eq. (39). When $N < 1$ we have no general argument for the behavior of $B_{22}$ in the massless limit. We may however restrict our attention to the universality class corresponding to the standard continuum limit, and within this class we may consider the specific lattice example of the minimal nearest-neighbor coupling. For this action one may show that for arbitrary $N$

$$B_{11} = \frac{I_{N/2}(N\beta)}{I_{N/2-1}(N\beta)}\,,$$
$$B_{22} = \frac{I_{N/2+1}(N\beta)}{I_{N/2-1}(N\beta)} = 1 - \frac{B_{11}}{\beta}\,. \tag{52}$$

When $N \geq 1$ $B_{11}$ is strictly smaller than one for all finite values of $\beta$ and only in the limit $\beta \to \infty$ the massless regime is attained, in which case $B_{22} \to 1$ as well, as expected from the general argument. However when $N < 1$ one may numerically check that a finite value $\beta_c$ exists such that $B_{11}(\beta_c) = 1$: as a consequence $B_{22}(\beta_c) = 1 - \frac{1}{\beta_c} \neq 1$. Since $B_{22}$ is strictly different from one in the massless limit, we get from Eq. (49) that within this universality class

$$\left(1 + \frac{2}{N}\right) \widehat{g}_r^* \;=\; 2(N+2) \tag{53}$$

for $N < 1$. Let's notice that the two solutions connect very smoothly to each other (the function and its first derivative have the same left and right limit) because of the double zero at $N = 1$ in the contribution that does not vanish when $N > 1$. Fig. 4 shows $\left(1 + \frac{2}{N}\right) \widehat{g}_r^*$ versus $N$.

This is very reminiscent of what is going to happen when $d = 2$: for small $N$ $\beta_c$ is finite and there is a domain of analiticity in $N$ for $\widehat{g}_r$ around $N = -2$, while for large $N$ $\beta_c$ is infinite (asymptotic freedom) and analiticity in $1/N$ is present. The two regimes seem to meet smoothly at $N = 2$.

As a further check of our results we may consider the $N = 0$ case. This is a very simple model of self-avoiding walks in one dimension. All computations are straightforward and one obtains, with the notations adopted here,

$$\lim_{N \to 0} \left(1 + \frac{2}{N}\right) \widehat{g}_r^* \;=\; 4\;, \tag{54}$$

in full agreement with our general formula. Also intermediate steps are reproduced, with the identification $B_{11} = \beta$, $B_{22} = 0$.

## IV. NON-LINEAR $\sigma$ MODELS IN HIGHER DIMENSIONS.

Lattice non-linear $\sigma$ models, which we may choose to describe in terms of the standard $O(N)$-invariant nearest-neighbor action

$$S_L \;=\; -N\beta \sum_{x,\mu} \vec{s}_x \cdot \vec{s}_{x+\mu} \tag{55}$$

subject to the constraint $\vec{s}_x^{\,2} = 1$, when considered on a $d$-dimensional lattice with $d < 4$ have a nontrivial critical point $\beta_c \leq \infty$ whose neighborhood (scaling region) is properly described by a renormalized continuum field theory. This theory is in turn nothing but the $\gamma \to 0$ limit of the standard $O(N)$-invariant scalar field theory (linear $\sigma$ model). We may therefore study the critical properties (and in particular the fixed-point value of the renormalized coupling) of the symmetric phase of the $O(N)$ model by exploring the region $\beta \to \beta_c$ of the lattice model.

The left-hand-side of Eq. (9) has a simple reinterpretation in terms quantities defined within the associated lattice spin model. Setting

$$\chi = \sum_x \langle\, \vec{s}_0 \cdot \vec{s}_x \,\rangle \;,$$

$$m_2 = \sum_x x^2 \langle\, \vec{s}_0 \cdot \vec{s}_x \,\rangle \;,$$

$$\xi^2 = \frac{m_2}{2d\chi} \;=\; \frac{1}{m_r^2} \;,$$

$$\chi_4 = \sum_{x,y,z} \langle\, \vec{s}_0 \cdot \vec{s}_x \; \vec{s}_y \cdot \vec{s}_z \,\rangle_c \;, \tag{56}$$

one can argue that the combination

$$\frac{\chi_4}{\chi^2 \xi^d} \tag{57}$$

should either admit a non-trivial limiting value or vanish with logarithmic deviations from scaling when the critical line is approached. This is essentially a consequence of the existence of an unique diverging relevant scale in the scaling region. It is furthermore trivial to show that in the scaling region, $m_r \to 0$,

$$f \;\equiv\; -N \frac{\chi_4}{\chi^2 \xi^d} \;=\; \left(1 + \frac{2}{N}\right) \widehat{g}_r \;, \tag{58}$$

and in particular

$$f^* \equiv f(\beta_c) \;=\; \left(1 + \frac{2}{N}\right) \widehat{g}_r^* \;, \tag{59}$$

We also mention that $f$ can be written in terms of the Binder cumulant defined on a $L^d$ lattice

$$U_L \;=\; 1 + \frac{2}{N} - \frac{\langle \left(\vec{S} \cdot \vec{S}\right)^2 \rangle}{\left(\langle \vec{S} \cdot \vec{S} \rangle\right)^2} \;, \tag{60}$$

where $\vec{S} = \sum_x \vec{s}_x$. Indeed

$$f \;=\; N \lim_{L \to \infty} U_L \left(\frac{L}{\xi_L}\right)^d \;. \tag{61}$$

As already mentioned in Sec. III, there is a crucial dependence on the space dimensionality as well as on $N$. In two dimensions it is well known that models with $-2 \leq N \leq 2$ are well described at criticality by conformal field theories with $c \leq 1$. In particular at $N = -2$ the fixed point is Gaussian, $N = 0$ corresponds to a model of self-avoiding random walks, $N = 1$ is the solvable Ising model, and $N = 2$ is the $XY$ model showing the Kosterlitz-Thouless critical phenomenon, characterized by an exponential singularity at a finite $\beta_c$.

When $-2 \leq N \leq 2$ the critical point occurs at a finite value of $\beta_c$, which should coincide with the convergence radius of the strong coupling series. When $N \geq 3$ there is apparently no criticality for any finite value of $\beta$. This is consistent with the Mermin-Wagner's theorem on the absence of spontaneous symmetry breakdown for two-dimensional continuous symmetry and with the weak coupling (large $\beta$) prediction of asymptotic freedom and dynamical mass generation for this class of models. Large-$N$ results and the $1/N$ expansion are completely consistent with the above picture. From the point of view of the renormalized coupling analysis it is however impossible to distinguish between the two behaviors, since they are both compatible with a non-zero value of $f^*$.

We now briefly present some large-$N$ results regarding $f$ and its limit at $\beta_c$. On the lattice, using the action (55), the large-$N$ limit of $f(\beta)$ can be easily obtained from the saddle-point equation

$$\beta \;=\; \int \frac{d^d q}{(2\pi)^d} \frac{1}{\widehat{q}^2 + m_0^2}\;, \tag{62}$$

where $\widehat{q}^2 \;\equiv\; 4\sum_\mu \sin^2(q_\mu/2)$, and the relation

$$f(\beta) \;=\; -\frac{2}{m_0^{4-d}\frac{\partial\beta}{\partial m_0^2}}\;. \tag{63}$$

In 2-d the above equations are made more explicit by writing

$$\beta = \frac{1}{2\pi} k K(k)\;,$$
$$f = 4\pi \frac{1+k}{kE(k)}\;, \tag{64}$$

where $k = \left(1+\frac{m_0^2}{4}\right)^{-1}$, $K$ and $E$ are elliptic functions. Fig. 5 shows $f(\beta)$ versus $\beta$. In the large-$\beta$ limit the continuum result (28), i.e. $f^* = 8\pi$, is recovered.

Our $1/N$ expansion analysis of Sec. II leads to the evaluation of the $O(1/N)$ correction to $f^*$, indeed in two dimensions:

$$f^* \;=\; 8\pi\left[1+\frac{f_1}{N}+O\left(\frac{1}{N^2}\right)\right] \tag{65}$$

with $f_1 = -0.602033....$

In three dimensions we face a quite different situation. The critical point occurs at a finite value $\beta_c$ for all values of $N$. At $N=\infty$ [11]

$$\beta_c(N=\infty) \;=\; \int \frac{d^3 q}{(2\pi)^3}\frac{1}{\widehat{q}^2} \;=\; 0.252731.... \tag{66}$$

The four-point renormalized coupling $f(\beta)$ at $N=\infty$ is shown in Fig. 6. At the critical point $f^* = 16\pi$. It is also possible to estimate the deviation of $\beta_c(N)$ from Eq. (66) by the $1/N$ expansion technique presented first in Ref. [11], leading to the relationship

$$\beta_c(N) \;=\; \beta_c(\infty) + \frac{b_1}{N} + O\left(\frac{1}{N^2}\right)\;, \tag{67}$$

where

$$
\begin{aligned}
b_1 &= -\int \frac{d^3q}{(2\pi)^3}\Delta^{(0)}(q)\int \frac{d^3p}{(2\pi)^3}\frac{1}{(\hat{p}^2)^2}\left[\frac{1}{2\widehat{(p+q)}^2}+\frac{1}{2\widehat{(p-q)}^2}-\frac{1}{\hat{q}^2}\right] , \\
\Delta^{(0)}(q)^{-1} &= \frac{1}{2}\int \frac{d^3r}{(2\pi)^3}\frac{1}{\hat{r}^2\widehat{(r+q)}^2} \; .
\end{aligned}
\tag{68}
$$

Numerically $b_1 \simeq -0.117$. It is also important to possess an estimate of the value of the internal energy $E$ at the criticality

$$
E_c(N) \;=\; E_c(\infty) + \frac{e_1}{N} + O\left(\frac{1}{N^2}\right) \; . \tag{69}
$$

We notice that

$$
\begin{aligned}
E &= \langle\, \vec{s}_x \cdot \vec{s}_{x+\mu}\,\rangle \;=\; \frac{1}{Nd}\frac{\partial}{\partial\beta}\ln Z(\beta,N) \\
&= 1-\frac{1}{2\beta d}+\frac{m_0^2}{2d}+\frac{1}{2Nd}\left[\frac{2}{\beta}+\bar{\Sigma}_1^{(b)}(m_0^2)\right]+O\left(\frac{1}{N^2}\right) ,
\end{aligned}
\tag{70}
$$

where $\bar{\Sigma}_1^{(b)}(m_0^2)$ is the lattice counterpart of $\Sigma_1^{(b)}(m_0^2)$, defined in Eq. (15). By setting $d=3$ and by considering the $m_r^2 \to 0$ limit we then obtain

$$
E_c(N) \;=\; 1-\frac{1}{6\beta_c(\infty)}+\frac{1}{6N}\left[\frac{b_1}{\beta_c(\infty)^2}+\frac{2}{\beta_c(\infty)}-\int\frac{d^3q}{(2\pi)^3}\frac{\Delta^{(0)}(q)}{\hat{q}^2}\right]+O\left(\frac{1}{N^2}\right) \; . \tag{71}
$$

Numerically we obtained $E_c(\infty) = 0.340537...$ and $e_1 \simeq -0.07$.

For the three dimensional case our $O(1/N)$ calculation of $f^*$ gives

$$
f^* \;=\; 16\pi\left[1+\frac{f_1}{N}+O\left(\frac{1}{N^2}\right)\right] \tag{72}
$$

with $f_1 = -1.54601....$

In the two and three dimensional $O(N)$ models a number of techniques have been applied to the determination of $f^*$. In particular we mention the $\phi^4$ field theoretical approach at fixed dimensions proposed by Parisi [12] and developed in Refs. [13,14], making also use of Borel resummation techniques (see for example Refs. [15,16] for a review on this approach). This method has been applied to $N=1$ (Ising models) in 2-d [14], $N=0,1,2,3$ [13,14,17] and many larger values of $N$ [17] in 3-d, leading to rather precise estimate of $f^*$, especially in 3-d. In order to compare our results with the field theoretical calculations we must keep in mind that it is customary to rescale the coupling in such a way that for all values of $N$ the one-loop fixed point value of the new coupling $\bar{g}$ be exactly one [13,14]. By comparing the one loop expression of the $\beta$-function, which in our notation would be

$$
\beta(\hat{g}_r) \;=\; (d-4)\hat{g}_r + \frac{N+8}{N}\,\frac{\Gamma\left(3-d/2\right)}{(4\pi)^{d/2}}\hat{g}_r^2 + O(\hat{g}_r^3) \, , \tag{73}
$$

we find

$$\bar{g}^* = \frac{N+8}{N+2}\frac{\Gamma(2-d/2)}{2(4\pi)^{d/2}} f^* . \tag{74}$$

We also mention a determination of $f^*$ for $N = 0$ in 3-d by working directly with the self-avoiding random walk model [18], which turns out to be in full agreement with the corresponding $\phi^4$ field theoretical calculation [13,14].

Estimates of $f^*$ can also be obtained by Monte Carlo simulations using the lattice formulation of the theory, by directly measuring $f(\beta)$. Numerical studies concerning the four-point coupling have been presented in the literature for some two-dimensional models: for $N = 1$ [19], and $N = 2, 3$ [20]. The comparison with these works must take into account the extra factor $N$ in our definition (61) of the four-point coupling $f$.

Finally let's briefly comment on the $d = 4$ case. In this case it is not possible to define a non-trivial limit for the non-linear $\sigma$ model in the strict $\gamma = 0$ regime, at least within the $1/N$ expansion, since we obtain the naive result $\Delta(k) = 0$ implying $\widehat{g}_r^* = 0$. This is however consistent with the common expectation that $O(N)$-invariant models in four dimensions may only have a trivial fixed point, in which case the critical region should be characterized by logarithmic deviations from scaling. That this is the case has been shown by Lüscher and Weisz by making use of the strong coupling expansion in a beautiful series of papers [1,3–6]. Kristjansen and Flyvbjerg [21] in turn have developed the lattice $1/N$ expansion of the $O(N)$ invariant models in four dimensions both in the symmetric and in the broken phase, finding substantial agreement with Refs. [1,3–6] at $N = 4$ in the region around criticality.

We may add that, by properly manipulating the expression presented in the Appendix in the limit $d \to 4$, it is possible to compute exactly (albeit only numerically) the $1/N$ correction to the $\beta$-function of the $(\vec{\phi}^2)^2$ model for all values of the (running) renormalized coupling. In turn these results might be used to improve our understanding of the non-perturbative limit of a strongly interacting Higgs sector on the line traced by Refs. [22,23], where the leading order result was analyzed.

## V. STRONG COUPLING ANALYSIS.

The non-triviality issue can be also investigated by high-temperature series methods formulating the theory on the lattice. We consider the nearest-neighbor formulation (55) of $O(N)$ vector models. Notice that in (55) we have introduced a rescaled inverse temperature $\beta$.

The strong coupling expansion of $f(\beta)$ has the following form

$$f(\beta) = \frac{1}{\beta^{d/2}}\left[2 + \sum_{i=1}^{\infty} a_i \beta^i\right] . \tag{75}$$

Series up to $14^{\rm th}$ order of the quantities involved in the definition of $f(\beta)$, i.e. $\chi$, $m_2$ and $\chi_4$, have been calculated by Lüscher and Weisz [1], and rielaborated by Butera et al. [2]. From such series one can obtain $A_d(\beta) \equiv \beta^{d/2} f(\beta)$ up to $13^{\rm th}$ order. We mention that, for other purposes, we have calculated the strong coupling series of the Green function $G(x) \equiv \langle\, \vec{s}_0 \cdot \vec{s}_x \,\rangle$ up to $21^{\rm th}$ order in 2-d, and up to $15^{\rm th}$ in 3-d, obtaining the corresponding series of the energy, the magnetic susceptibility and the second moment correlation length to the same order [24].

We also considered strong coupling series in the energy $f(E)$, which can be obtained by inverting the strong coupling series of the energy $E = \beta + O(\beta^3)$ and substituting in Eq. (75):

$$f(E) = \frac{1}{E^{d/2}} \left[ 2 + \sum_{i=1}^{\infty} e_i E^i \right] . \tag{76}$$

From our strong coupling series of the energy [24], we could calculate $B_d(E) \equiv E^{d/2} f(E)$ up to $13^{\rm th}$ order.

Before describing our analysis of the above series based on Padè approximants (PA's) technique (see Ref. [25] for a review on the analysis of strong coupling series), we recall that PA's are expected to converge well to meromorphic analytic functions. More flexibility is achieved by applying the PA analysis to the logarithmic derivative of the strong coupling series considered (Dlog-PA analysis), and therefore enlarging the class of functions which can be reproduced to those having branch-point singularities. In general more complicated structures may arise, such as confluent singularities, which are sources of systematic errors for a PA (or Dlog-PA) analysis. In particular confluent singularities at $\beta_c$, i.e. confluent corrections to scaling arising from irrelevant operators [14,26], lead in general to a non-diverging singularity of $f(\beta)$ at $\beta_c$. Indeed in the presence of confluent singularities we would expect $f(\beta)$ to behave as

$$f(\beta) \simeq f^* + c(\beta_c - \beta)^{\Delta} \tag{77}$$

close to $\beta_c$ with $\Delta > 0$. Such a behavior close to $\beta_c$ cannot be reproduced by a PA's or Dlog-PA's, while it could be detected by a first or higher order differential approximant analysis [27]. Therefore in order to reduce systematic errors one should turn to more general and flexible analysis, such as differential approximants, which, on the other hand, require many terms of the series to give stable results. We tried also this type of analysis without getting stable and therefore acceptable results, very likely due to the relative shortness of the available series. We then expect PA and Dlog-PA analysis to be subject to larger systematic errors when confluent singularities are more relevant, as in 3-d models at small $N$, where they represent a serious problem also in the determination of the critical exponents from the available strong coupling series.

It is important to notice that the accuracy and the convergence of the PA estimates may change when considering different representations of the same quantity, according to how well the function at hand can be reproduced by a meromorphic analytic function. By comparing the results from different series representations of the same quantity one may check for possible systematic errors in the resummation procedure employed. To this purpose, in our study we will compare estimates of $f^*$ coming from the strong coupling series of both $f(\beta)$ and $f(E)$. Exact results at $N = \infty$ presented in the previous section, beside giving an idea of the behavior of $f(\beta)$ at finite $N$, represent useful benchmarks for strong coupling methods.

The most direct way to evaluate $f^* \equiv f(\beta_c)$ would consist in computing $[l/m]$ PA's $A_{l/m}(\beta)$ from the available series of $A_d(\beta)$, and evaluating $\beta^{-d/2} A_{l/m}(\beta)$ at the critical point $\beta_c$ (at least if $\beta_c < \infty$; if $\beta_c = \infty$ things are trickier as we will discuss below). This simple procedure works already reasonably well, but we found more effective a Dlog-PA analysis, which showed a greater stability and whose results will be presented in the following.

Our Dlog-PA analysis consisted in computing $[l/m]$ PA to the strong coupling series of the logarithmic derivative of $A_d(\beta)$, let's indicate them with $\text{Dlog}_{l/m}A_d(\beta)$, and then a set of corresponding approximants $f_{l/m}(\beta)$ to $f(\beta)$, which are obtained by reconstructing $f(\beta)$ from the logarithmic derivative of $A_d(\beta)$:

$$f_{l/m}(\beta) \;=\; \frac{2}{\beta^{d/2}} \exp \int_0^\beta d\beta' \, \text{Dlog}_{l/m} A_d(\beta') \;, \tag{78}$$

All the approximants with

$$l + m \;\geq\; 10 \;, \qquad m \geq l \geq 4 \;. \tag{79}$$

were considered in order to check the stability of the procedure. Notice that, for given $l, m$, the number of terms of the series of $A_d(\beta)$ used by the corresponding Dlog-PA is $n = l+m+1$. Once the approximant $f_{l/m}(\beta)$ is computed, if $\beta_c$ is finite, its value at $\beta_c$ provides an estimate of $f^*$.

This requires a rather precise determination of $\beta_c$, which is in some cases available in the literature from strong coupling and numerical Monte Carlo studies. When $\beta_c$ was not known, we estimated it from a Dlog-Padè analysis of the strong coupling series of the magnetic susceptibility (up to $21^{\text{th}}$ order in 2-d and $15^{\text{th}}$ in 3-d). Our strong coupling determinations of $\beta_c$ in 3-d models at large $N$ compare very well with the $O(1/N)$ calculation (67), as shown in Fig. 7. Let's notice that the error on the value of $\beta$ is small enough not to be relevant for the estimate of $f^*$.

In order to better understand the analytic structure of $f(\beta)$ we have done a detailed study of the complex-plane singularities of the Dlog-PA's of $A_d(\beta)$. We have first checked hyperscaling. A violation of hyperscaling would lead to a behavior $f(\beta) \sim A_d(\beta) \sim (\beta_c - \beta)^\epsilon$ for $\beta \to \beta_c$, and thus the Dlog-PA's would show a simple pole at $\beta = \beta_c$. We recall that a Dlog-PA analysis is in general very efficient in detecting power-law singularities. We have found no evidence of such a pole, confirming hyperscaling arguments. However notice that Eq. (77) when $1 > \Delta > 0$ implies a behavior

$$\text{Dlog} A_d(\beta) \;\sim\; (\beta_c - \beta)^{\Delta - 1} \tag{80}$$

close to $\beta_c$. In two dimensions $\Delta \simeq 1$ and therefore we do not expect to find singularities around $\beta_c$. This is confirmed by the analysis of $A_2(\beta)$. In three dimensions instead, at least for small $N$, $\Delta \simeq 0.5$ is expected, thus $A_3(\beta)$ should behave as in Eq. (80), and in the Dlog-PA's the singularity should be mimicked by a shifted pole at a $\beta$ larger than $\beta_c$. Indeed in the analysis of $A_3(\beta)$ we have found a singularity typically at $\beta \simeq 1.1 \div 1.2\ \beta_c$. This fact will eventually affect the determination of $f(\beta)$ close to $\beta_c$ by a systematic error. However since the singularity is integrable the error must be finite, and the analysis shows that such errors are actually reasonably small.

Sometime PA's showed spurious singularities on the positive real axis (or very close to it) for $\beta \lesssim \beta_c$. We considered these approximants defective, and discarded them from the analysis. Such defective PA's were a minority, as the Tables show. The only stable singularity detected by the Dlog-PA's of $A_d(\beta)$ lies in the negative $\beta$ axis and closer to the origin than $\beta_c$: it turns out to be nothing but a regular zero of $A_d(\beta)$. The position of this negative zero is reported in the Tables I,II,III and V for several values of $N$.

As final estimates of $f^*$, reported in the Tables IV and VI, we take the average of the values $f_{l/m}(\beta_c)$ from the non-defective PA's using all available terms of the series, i.e. those with $n = l+m+1 = 13$. The errors displayed in the Tables IV and VI are just indicative, they are the variance around the estimate of $f^*$ of the results coming from all PA's considered (cfr. (79)), which should give an idea of the spread of the results coming from different PA's. Such errors do not always provide a reliable estimate of the systematic errors, which may be underestimated especially when the structure of the function (or of its logarithmic derivative) is not well approximated by a meromorphic analytic function. In such cases a more reliable estimate of the systematic error would come from the comparison of results from the analysis of different series representing the same quantity, which in general are not expected to have the same structure.

For this reason we have considered the series in the energy variable, which we have analyzed exactly as $f(\beta)$. In this case instead of $\beta_c$ we needed $E_c$, the energy at the critical point. When the value of $E_c$ was not available in the literature, we estimated it by the first real positive singularity found in the analysis of the available strong coupling series of the magnetic susceptibility expressed in powers of $E$. This procedure provides an estimate of $E_c$ much less precise than $\beta_c$ (see Table V for the values obtained in 3-d), but sufficiently good to our purposes, given that $f(E)$ is smooth around $E_c$. In Fig. 8 we compare our determinations of $E_c$ in 3-d models with the large-$N$ result (71), showing agreement within the uncertainty of the strong coupling results.

In asymptotically free models where $\beta_c = \infty$, the task of determining $f^*$ from a strong coupling approach appears much harder. On the other hand, since at sufficiently large $\beta$ we expect that

$$|f(\beta) - f^*| \sim \xi^{-2}, \tag{81}$$

a reasonable estimate of $f^*$ could be obtained at $\beta$-values corresponding to large correlation length, $\xi \gtrsim 10$ say, where the curve $f(\beta)$ should be already stable (scaling region). Notice that this is the same idea underlying numerical Monte Carlo studies. Another interesting possibility is to change variable from $\beta$ to the energy $E$, and analyze the series in powers of $E$. In the energy variable the continuum limit is reached for $E \to 1$, and therefore the strong coupling approach to the continuum limit appears more feasible. In order to reach the continuum limit from strong coupling, we believe this change of variable to be effective especially for the analysis of dimensionless ratios of physical quantities.

To begin with we present the results obtained for the 2-d models. PA's for $-1 \leq N < 2$ are quite stable, giving estimates of $f^*$ very close to each other as shown in Table I, where the values of the approximants $f_{l/m}(\beta)$ at $\beta_c$ are reported. For $N < -1$, due to the instability of the corresponding PA's, we could not get reliable estimates of $f^*$. Fig. 9 shows $f_{6/6}(\beta)$ versus $\beta/\beta_c$ for various values of $N$. Differences in the other PA's were of the order of the width of the lines drawn in the figure. Final estimates of $f^*$ are reported in Table IV. In order to check possible systematic errors in our analysis, we applied the above procedure to the $13^{\text{th}}$ order series of $B_2(E)$ of the Ising model ($N = 1$). The value of $f(E)$ at $E_c = \sqrt{2}/2$, the energy value at $\beta_c$, must give again $f^*$. As in the analysis of $A_2(\beta)$, approximants $f_{l/m}(E)$ turn out to be rather stable, leading to the result $f^* = 14.64(10)$, which is perfectly consistent with the estimate $f^* = 14.63(7)$ coming from the analysis of $A_2(\beta)$.

For the Ising model ($N = 1$) our estimate of $f^*$ is in agreement with the result of Ref. [28], obtained by a slightly different strong coupling analysis (the value reported there is $f^* = 14.67(5)$), and with the estimates by $\phi^4$ field theory calculations at fixed dimensions [14] and numerical Monte Carlo simulations [19] (see Table IV).

The 2-d $XY$ model ($N = 2$) is expected to follow the pattern of a Kosterlitz-Thouless critical phenomenon, whose critical region is characterized by a correlation length diverging exponentially with respect to $\tau \equiv \beta_c - \beta$: $\xi \sim \exp(b/\tau^\sigma)$ with $\sigma = 1/2$. For this model the values of $f_{l/m}(\beta)$ and $f_{l/m}(E)$ respectively at $\beta_c = 0.559(3)$ [29,30] and $E_c = 0.722(3)$ [29] are reported in Table II. Fig. 10 shows various non-defective $f_{l/m}(\beta)$ at $N = 2$, comparing them with the Monte Carlo results obtained recently by Kim [20] for correlation lengths: $5 \lesssim \xi \lesssim 70$. The agreement is very good especially for PA's obtained using all available 13 terms of the strong coupling series. PA's of the series in $E$ turn out to be more stable, as shown in Table II and in Fig. 11, giving a perfectly consistent result for $f^*$. Our final estimate is $f^* = 18.2(2)$ which is slightly larger than the Monte Carlo result $f^* = 17.7(2)$ [20] (this number has been obtained by taking only data for $\xi \gtrsim 25$ of Ref. [20] and taking into account the extra factor $N$ in our definition (58)), but definitely consistent.

When $N \geq 3$ the critical point moves to infinity making the determination of $f^*$ from strong coupling harder. For such models our analysis should be considered just exploratory due to the shortness of the available series, but as we will see the results look promising. In order to give an idea of the stability of our resummation procedure in this case, in Table III we report the values taken by $f_{l/m}(\beta)$ and $f_{l/m}(E)$ for couples of $\beta$ and $E$ corresponding to a correlation length $\xi \simeq 10$. In Fig. 12 various $f_{l/m}(\beta)$ at $N = 3$ are drawn and compared with the Monte Carlo results of Ref. [20], obtained for correlation lengths $10 \lesssim \xi \lesssim 120$. The curves corresponding to different $f_{l/m}(\beta)$ are very close up to $\beta \simeq 0.5$ (see also Table III). At $\beta \gtrsim 0.5$ we observe that curves from different PA's become more and more stable with increasing $n = l + m + 1$, improving the agreement with the Monte Carlo data. Anyway, the agreement is quite good, even for $\beta \simeq 0.6$, corresponding to $\xi \simeq 100$. Fig. 13 shows some approximants $f_{l/m}(E)$ computed from the series in the energy. At $E = 1$ they give consistent results within 5-10%. Similar results are observed for $N = 4$, as shown in Fig. 14, where some $f_{l/m}(\beta)$ are plotted. Notice that at $\beta = 0.6$ $\xi \simeq 25$ [31].

In order to get an estimate of $f^*$ for both $N = 3$ and $N = 4$ we considered the values of $f_{l/m}(\beta)$ and $f_{l/m}(E)$ at the largest values of $\beta$ and $E$ where they are still stable, i.e. $\beta \simeq 0.5$ and $E \simeq 0.6$, which correspond to an acceptably large correlation length $\xi \simeq 10$. So, from data in Table III, our final estimate at $N = 3$ is $g^* \simeq 19.8$, with an uncertanty of few per cent, which compares very well with the Monte Carlo result $f^* = 19.6(2)$ (obtained by fitting all data reported in Ref. [20] to a constant) and $O(1/N)$ calculation $f^* \simeq 20.09$. At $N = 4$ our estimate is $f^* \simeq 21.2$ against $f^* \simeq 21.35$ coming from the $O(1/N)$ calculation.

Fig. 15 summarizes our 2-d results: it shows our strong coupling estimates of $f^*$ versus $N$, comparing them with the available estimates of $f^*$ from alternative approaches: $\phi^4$ field theory, Monte Carlo and $1/N$ expansion techniques. There is a general agreement, in particular the $O(1/N)$ calculation $f^* \simeq 8\pi\left(1-\frac{0.602033}{N}\right)$ fits very well data down to $N=3$. Furthermore we observe the linear approach of $f^*$ toward zero for $N \to -2$, similarly to the $d=1$ case.

Let's now consider 3-d $O(N)$ $\sigma$ models, which present a critical behavior at a finite $\beta$ for all values af $N$. In order to check possible systematic errors we analyzed both the strong coupling series of $f(\beta)$ and $f(E)$. Table V shows a summary of the estimates of $f^*$ from the values of the approximants $f_{l/m}(\beta)$ at $\beta_c$ and $f_{l/m}(E)$ at $E_c$. Final estimates of $f^*$ from the analysis of the strong coupling series in $\beta$ and in $E$ are reported in Table VI. We recall that the errors diplayed in the Table VI are related to the spread of the PA results, while an estimate of the true systematic errors could only come from the comparison of results from different series associated to the same quantity.

Fig. 16 shows typical curves of $f(\beta)$ obtained by [6/6] PA's (sometimes we used [7/5] PA's when the [6/6] ones were defective). The error bars displayed at $\beta/\beta_c = 1$ show the spread of the estimates of $f^*$ from different non-defective $[l/m]$ PA's.

At large $N$, typically $N \geq 3$, both series of $f(\beta)$ and $f(E)$ give consistent results, which should be an indication of small systematic errors. As further check of our resummation procedure in the large-$N$ region, we repeated our analysis at $N=\infty$. We found that most of the approximants $f_{l/m}(\beta)$ constructed from the 13$^{\rm th}$ order series of $A_3(\beta)$, if plotted in Fig. 6, would not be distinguishable from the exact curve. The analysis of the $N=\infty$ 13$^{\rm th}$ order series of $A_3(\beta)$ and $B_3(E)$ would have given respectively $f^* = 50.25(6)$ and $f^* = 50.27(6)$ against the exact value $f^* = 16\pi = 50.2654....$ Therefore everything seems to work fine at large $N$. On the contrary, at small $N$ there are discrepancies between the analysis in $\beta$ and in $E$, which are definitely larger than the typical spread of the PA estimates of $f^*$ from each series. Such differences give somehow an idea of the size of the systematic errors of our analysis when applied to these values of $N$.

In Table VI for comparison we give also the results from $\phi^4$ field theory and $1/N$ expansion. Fig. 17 summarizes all available results for $f^*$. There as a strong coupling estimate of $f^*$ we show the average of the results from the series $f(\beta)$ and $f(E)$, while their difference is used as an estimate of the systematic error.

At large $N$, $N \geq 8$, there is a substantial general agreement: estimates from the strong coupling approach, $O(1/N)$ calculation $f^* \simeq 16\pi\left(1-\frac{1.54601}{N}\right)$, and $\phi^4$ field theory differ at most by 1% to each other. At small $N$, $N = 0,1,2$, our strong coupling estimates show relevant discrepancies with the field theoretical calculations, which are of the size of the differences between the results coming from the analysis of $f(\beta)$ and $f(E)$, and therefore they should be caused by systematic errors in the strong coupling analysis employed. Anyway such discrepancies are not dramatic, indeed they are at most 5% and decrease with increasing $N$.

In conclusion we have seen that, in two and three dimensions 13 terms of the strong coupling series of $A_d(\beta)$ and $B_d(E)$ are already sufficient to give quite stable results, which compare very well with calculations from other techniques, such as $\phi^4$ field theory at fixed dimensions, Monte Carlo simulations and $1/N$ expansion. Of course an extension of the series of $f$ would be welcome, especially for two reasons:

(i) to further stabilize the PA's in the asymptotically free models and obtain reliable estimates at values of $\beta$ corresponding to large correlation lengths $\xi \gtrsim 100$, and moreover check if the change of variable $\beta \to E$ allows one to get a reliable strong coupling estimate of $f^*$ in the continuum limit $E \to 1$;

(ii) to see if the apparent discrepancies at small $N$ in 3-d with the more precise $\phi^4$ field theory calculations get reduced.

An extension of the series of $f(\beta)$ may also allow more accurate and flexible analysis, like differential approximants, which in general require many terms of the series in order to give stable results, and which could provide a better reconstruction of $f(\beta)$ from its strong coupling series, taking properly into account the confluent singularities, which should be the major source of systematic error in 3-d models at small $N$.

## VI. CONCLUSIONS

We computed the dependence of the renormalized four-point coupling $g_r$ from the renormalized mass $m_r$ and the bare coupling to $O(1/N)$ for $O(N)$-invariant $(\vec{\phi}^2)^2_d$ theories ($d \leq 4$) in the symmetric phase. As a consequence we obtained expressions for the $\beta$-function and its fixed point $g_r^*$ within the same approximation.

We extracted an independent determination of $g_r^*$ from the strong coupling analysis of the $O(N)$ non-linear $\sigma$ models, which we performed for $d = 2, 3$ and selected values of $N$ in the whole range $N > -2$, applying resummation techniques both in the inverse temperature variable $\beta$ and in the energy variable $E$. In two dimensions and for $N$ sufficiently large ($N \geq 3$) in three dimensions we found a good agreement with the $\phi^4$ fixed-dimension field theory estimates, and we could also check consistency with the $1/N$ prediction, thus seemingly indicating good convergence properties of the $1/N$ expansion at least when applied to the above quantities. In three dimensions and for small $N$, however, some discrepancy between resummations of the series in $\beta$ and in $E$ occurred, which we interpreted as an indication of systematic errors, and which was also reflected into a small disagreement with results presented in the literature and obtained with other techniques, like $\phi^4$ field theory at fixed dimensions. Such discrepancy might be significantly reduced by knowing a few more terms in the strong coupling series, whose feasibility seems to be well within the range of present day strong coupling techniques. In our opinion improving the strong coupling analysis might lead to a determination of the fixed point value of the renormalized four-point coupling with a precision comparable to, or even better than, the best available results. We stress the crucial role played by the comparison of series in the variables $\beta$ and $E$ in order to estimate the relevance of systematic errors.

## ACKNOWLEDGMENTS

It is a pleasure to thank A. J. Guttmann and A. D. Sokal for useful and stimulating discussions.

## APPENDIX A

By applying Eq. (35) to Eq. (24) and making explicit use of Eq. (36), we may obtain the following explicit representation of the $O(1/N)$ contribution to the $\beta$-function of $O(N)$ models in $d$ dimensions:

$$\begin{aligned}
\frac{\beta^{(1)}(\widehat{g}_r)}{\widehat{g}_r^2} &= (d-3)2^{d-1}\beta_0 + \frac{2}{d}(d-1)^2\left(d-4+\beta_0\widehat{g}_r\right)^2 \int \frac{d^d u}{(2\pi)^d} \frac{1}{\left[1+\widehat{g}_r\Pi(u)\right]^2} \frac{1}{\left(4+u^2\right)^2} \\
&+2\int \frac{d^d u}{(2\pi)^d}\left[\frac{\beta_0\widehat{g}_r+d-4}{(1+\widehat{g}_r\Pi(u))^2} - \frac{\beta_0\widehat{g}_r}{1+\widehat{g}_r\Pi(u)}\right]\left[\frac{1}{(1+u^2)^3} + \frac{3}{(1+u^2)(4+u^2)}\left(\frac{\frac{d}{4}-2}{1+u^2} - \frac{d-1}{4+u^2}\right)\right] \\
&-\frac{2}{d}\int \frac{d^d u}{(2\pi)^d}\frac{(\beta_0\widehat{g}_r+d-4)^2}{(1+\widehat{g}_r\Pi(u))^2}\left[\frac{1}{(1+u^2)^3} + \frac{3}{(1+u^2)(4+u^2)}\left(\frac{\frac{d}{4}-1}{1+u^2} + \frac{d-1}{4+u^2}\right)\right] \\
&-4\int \frac{d^d u}{(2\pi)^d}\left[\frac{\beta_0\widehat{g}_r+d-4}{(1+\widehat{g}_r\Pi(u))^3} - \frac{\beta_0\widehat{g}_r+\frac{d}{2}-2}{(1+\widehat{g}_r\Pi(u))^2}\right]\frac{1}{(4+u^2)^2}\left(\beta_0\widehat{g}_r+d-4-\frac{3}{1+u^2}\right)^2 \\
&-4\int \frac{d^d u}{(2\pi)^d}\frac{\beta_0\widehat{g}_r(\beta_0\widehat{g}_r+d-4)}{(1+\widehat{g}_r\Pi(u))^2}\frac{1}{(4+u^2)^2}\left(\beta_0\widehat{g}_r+d-4-\frac{3}{1+u^2}\right)\,,
\end{aligned} \tag{A1}$$

where we have introduced the rescaled integration variable $u \equiv k/m_r$. By noticing that, according to its definition (37)

$$0 \geq \Pi(u) \geq -\frac{1}{g_*}\,, \tag{A2}$$

it is easy to get convinced that all integrals appearing in Eq. (A1) are well-defined and finite as long as $d \leq 4$ and $\widehat{g}_r < g_*$. Moreover it is possible to perform a series-expansion of Eq. (A1) in the powers of $\widehat{g}_r$, reproducing order by order standard perturbation theory results [17], and in particular to leading order:

$$\frac{\beta^{(1)}(\widehat{g}_r)}{\widehat{g}_r^2} \underset{\widehat{g}_r\to 0}{\longrightarrow} 8\beta_0 \tag{A3}$$

for all values of $d$.

For the sake of comparison in Figs. 18 and 19 we plot the function $\beta^{(1)}(\bar{g})$, where $\bar{g}$ has been defined as in Refs. [13,14,17] such that $\bar{g}^* = 1$ at $N = \infty$ (see also Sec. IV and Eq. (74)), respectively for $d = 3$ and $d = 2$.

We recall once more that

$$\bar{g} = \frac{N+8}{N}\frac{\beta_0}{4-d}\widehat{g}\,, \tag{A4}$$

and by definition we set

$$\beta^{(1)}(\bar{g}) = (4-d)\sum_{n=1}^{\infty}\beta_n\bar{g}^{n+2}\,. \tag{A5}$$

In Table VII we report all values in $d = 1, 2, 3$ such that $\beta_n \gtrsim 10^{-3}$. As a check of accuracy of the perturbative expansion we may employ the identity

$$6 + \sum_{n=1}^{\infty} \beta_n \; = \; -f_1 \; , \tag{A6}$$

where $f_1$ was defined and evaluated in Sec. IV (cfr. Eqs. (65) and (72)). Notice that for $d = 3$ the coefficients $\beta_n$ for $n \leq 5$ can also be extracted from the literature [17], and Eq. (A6) is already satisfied within 1% precision by the six-loop $\beta$-function. We mention that, in the case $d = 1$, $\beta^{(1)}(\bar{g})$ may actually be computed analytically, and the result is

$$\beta^{(1)}(\bar{g}) \; = \; \frac{3\bar{g}^2(1-\bar{g})^{3/2}}{4(3+\bar{g})^4} \left(648 - 3732\bar{g} + 5512\bar{g}^2 - 2183\bar{g}^3 - 330\bar{g}^4 - 27\bar{g}^5\right)$$
$$- \frac{6\bar{g}^2}{(3+\bar{g})^4} \left(81 - 6\bar{g} + 1750\bar{g}^2 - 1598\bar{g}^3 + 509\bar{g}^4\right) \; . \tag{A7}$$

The definitions (A4) and (A5) are obviously inappropriate in the limit $d \to 4$, in which case one may verify that

$$\frac{\beta^{(1)}(\widehat{g}_r)}{\widehat{g}_r^2} \; \longrightarrow \; 8\beta_0 \widehat{g}_r^2 - 9\beta_0^2 \widehat{g}_r^3 + O(\widehat{g}_r^4) \tag{A8}$$

where $\beta_0 \to \frac{1}{16\pi^2}$. Eq. (A8) in turn can be compared to the known perturbative evaluation around $d = 4$:

$$\beta(\widehat{g}_r) \; = \; (d-4)\widehat{g}_r + \frac{N+8}{N}\beta_0 \widehat{g}_r^2 - \frac{3(3N+14)}{N^2}\beta_0^2 \widehat{g}_r^3 + O(\widehat{g}_r^4) \; , \tag{A9}$$

finding complete agreement to $O(1/N)$.

It is conceivable to reinterpret the $d \to 4$ limit of Eq. (A1) in a nonperturbative sense by a principal-part prescription for the singularity occurring at the Landau pole $\bar{u}$ identified by the condition

$$\Pi(\bar{u}) \; = \; -\frac{1}{\widehat{g}_r} \; . \tag{A10}$$

Work in this direction is in progress.

## FIGURES

FIG. 1. Feynman rules for the $1/N$ expansion.

FIG. 2. Graphical definition of fundamental integrals.

FIG. 3. Identities among Feynman graphs.

FIG. 4. $f^*$ vs. $N$ for 1-d $O(N)$ models.

FIG. 5. $f(\beta)$ vs. $\beta$ for 2-d $O(\infty)$ model. The dotted horizontal line represents the continuum value $f^* = 8\pi$.

FIG. 6. $f(\beta)$ vs. $\beta$ for 3-d $O(\infty)$ model. The dotted horizontal line represents the continuum value $f^* = 16\pi$. The dashed vertical line indicates the critical point $\beta_c = 0.252631....$

FIG. 7. $\Delta\beta_c \equiv \beta_c(\infty) - \beta_c(N)$ versus $1/N$ in 3-d models, as estimated by a strong coupling analysis. The dashed line represents the $O(1/N)$ calculation (cfr. Eq. (67)).

FIG. 8. $\Delta E_c \equiv E_c(\infty) - E_c(N)$ versus $1/N$ in 3-d models, as estimated by a strong coupling analysis. The dashed line represents the $O(1/N)$ calculation (cfr. Eq. (69)).

FIG. 9. $f_{6/6}(\beta)$ vs. $\beta/\beta_c$ for various values of $N < 2$ in two-dimensional models.

FIG. 10. Some $f_{l/m}(\beta)$ are plotted versus $\beta$ for the 2-d $XY$ model ($N = 2$). The vertical dotted lines indicate the critical point: $\beta_c = 0.559(3)$ [29,30]. Monte Carlo data from Ref. [20] are also shown.

FIG. 11. Some $f_{l/m}(E)$ are plotted versus $E$ for the 2-d $XY$ model ($N = 2$). The vertical dotted lines indicate the value of the energy at the critical point: $E_c = 0.722(3)$, estimated from Monte Carlo data [29]. Monte Carlo data from Ref. [20] are also shown.

FIG. 12. Some $f_{l/m}(\beta)$ are plotted versus $\beta$ for the 2-d $O(3)$ model. Monte Carlo data from Ref. [20] are also shown.

FIG. 13. Some $f_{l/m}(E)$ are plotted versus $E$ for the 2-d $O(3)$ model. Monte Carlo data from Ref. [20] are also shown.

FIG. 14. Some $f_{l/m}(\beta)$ are plotted versus $\beta$ for the 2-d $O(4)$ model.

FIG. 15. For 2-d models we plot $f^*$ vs. $N$ obtained from our strong coupling analysis. For comparison field theoretical and Monte Carlo estimates are also shown. The dashed line represents the $O(1/N)$ calculation of $g^*$.

FIG. 16. $f(\beta)$ vs. $\beta/\beta_c$ for various values of $N$ in three dimensional models as obtained by a [6/6] PA (or [7/5] when the [6/6] one was defective). Error bars at $\beta/\beta_c = 1$ show the spread in the determination of $f^*$ from all PA's considered.

FIG. 17. $f^*$ vs. $N$ from our strong coupling analysis in 3-d. For comparison field theoretical estimates are also shown. The dashed line represents the $O(1/N)$ calculation of $f^*$. The dotted line indicates the value of $f^*$ at $N = \infty$.

FIG. 18. We plot $\beta^{(1)}(\bar{g})$ vs. $\bar{g}$ for the three-dimensional case.

FIG. 19. We plot $\beta^{(1)}(\bar{g})$ vs. $\bar{g}$ for the two-dimensional case.

# TABLES

TABLE I. For some 2-d $O(N)$ $\sigma$ models with $N < 2$ we report: the critical point, $\beta_c$; the estimate of the singularity of the PA's closest to the origin, $\beta_0$, which corresponds to a regular zero of $A_2(\beta)$; the values of the approximants $f_{l/m}(\beta)$ at the critical point. Asterisks mark defective PA's.

| $N$ | $\beta_c$ | $\beta_0$ | 5/5 | 4/6 | 5/6 | 4/7 | 6/6 | 5/7 | 4/8 |
|---|---|---|---|---|---|---|---|---|---|
| -1 | 0.3145(1) | -0.1315 | 5.27 | 5.32 | 5.27 | 5.28 | 5.27 | 5.31 | * |
| -1/2 | 0.3492(1) | -0.1506 | 7.87 | * | * | 8.04 | 7.85 | 8.10 | 8.00 |
| 0 | 0.379052(1) [33] | -0.1653 | 10.51 | * | * | 10.43 | 10.54 | 10.48 | * |
| 1/2 | 0.408545(8) | -0.1774 | 12.60 | 12.72 | 12.68 | * | 12.66 | 12.61 | 12.63 |
| 1 | 0.4406867... | -0.1878 | 14.65 | 14.72 | 14.70 | 14.67 | 14.69 | * | 14.57 |
| 3/2 | 0.4804(1) | -0.1969 | 16.76 | 16.76 | 16.76 | 16.60 | 16.83 | 16.83 | 16.47 |

TABLE II. For the 2-d $XY$ model ($N = 2$) we give some details on the analysis of the series of $f(\beta)$ (first line) and $f(E)$ (second line). We report: the critical point, $\beta_c$ ($E_c$); the estimate of the regular zero of $A_2(\beta)$ ($B_2(E)$) closest to the origin, $\beta_0$ ($E_0$); the values of the approximants $f_{l/m}(\beta)$ ($f_{l/m}(E)$) at $\beta_c$ ($E_c$). Asterisks mark defective PA's.

| $N$ | | | 5/5 | 4/6 | 5/6 | 4/7 | 6/6 | 5/7 | 4/8 |
|---|---|---|---|---|---|---|---|---|---|
| 2 | $\beta_c$ =0.559(3) [29,30] | $\beta_0$ =-0.2049 | 19.27 | 19.44 | * | 18.71 | * | * | 18.24 |
| | $E_c$ =0.722(3) [29] | $E_0$ =-0.2179 | 18.28 | 18.46 | 18.30 | 18.35 | 18.29 | * | 18.17 |

TABLE III. We give some details of the strong coupling analysis of the series $f(\beta)$ (first line) and $f(E)$ (second line) for two asymptotically free models: $N = 3, 4$. We report: the estimate of the regular zero of $A_2(\beta)$ ($B_2(E)$) closest to the origin, $\beta_0$ ($E_0$); the values of the approximants $f_{l/m}(\beta)$ ($f_{l/m}(E)$) at a value $\bar{\beta}$ ($\bar{E}$) corresponding to a correlation length $\xi \simeq 10$. The values of $E$ and $\xi$ are taken from Ref. [32] for $N = 3$, and Ref. [31] for $N = 4$. Asterisks mark defective PA's, i.e. PA's having singularities for $\beta \lesssim \bar{\beta}$ ($E \lesssim \bar{E}$ ).

| $N$ | | | $\xi$ | 5/5 | 4/6 | 5/6 | 4/7 | 6/6 | 5/7 | 4/8 |
|---|---|---|---|---|---|---|---|---|---|---|
| 3 | $\beta_0$ =-0.2188 | $\beta$ =0.5 | 11.05(1) | 20.3 | 20.6 | * | 20.0 | * | 19.5 | 19.8 |
| | $E_0$ =-0.2330 | $E$ =0.60157 | 11.05(1) | 19.9 | 20.0 | 19.9 | 19.9 | 19.9 | 19.9 | 19.8 |
| 4 | $\beta_0$ =-0.2305 | $\beta$ =0.525 | 10.32(3) | 21.8 | 22.4 | * | 21.3 | * | 20.6 | 21.0 |
| | $E_0$ =-0.2456 | $E$ =0.60089 | 10.32(3) | 21.2 | 21.3 | 21.3 | 21.3 | 21.2 | 21.4 | 21.1 |

TABLE IV. For 2-d $O(N)$ $\sigma$ models we report: the critical point $\beta_c$; $f^*$ from our strong coupling analysis, $f^*_{\rm sc}$; the $O(1/N)$ calculation of $f^*$, $f^*_{1/N}$; $f^*$ from $\phi^4$ field theory at fixed dimensions, $f^*_{\rm ft}$; $f^*$ from Monte Carlo simulations, $f^*_{\rm mc}$.

| $N$ | $\beta_c$ | $f^*_{\rm sc}$ | $f^*_{1/N}$ | $f^*_{\rm ft}$ | $f^*_{\rm mc}$ |
|---|---|---|---|---|---|
| -1 | 0.3145(1) | 5.29(3) | | | |
| -1/2 | 0.3492(1) | 8.0(1) | | | |
| 0 | 0.379052(1) [33] | 10.51(5) | | | |
| 1/2 | 0.408545(8) | 12.63(5) | | | |
| 1 | 0.4406867... | 14.63(7) | | 15.5(8) [14] | 14.3(1.0) [19] |
| 3/2 | 0.4804(1) | 16.7(2) | | | |
| 2 | 0.559(3) [29,30] | 18.2(2) | | | 17.7(2) [20] |
| 3 | $\infty$ | 19.8(4) | 20.09 | | 19.6(2) [20] |
| 4 | $\infty$ | 21.2(5) | 21.35 | | |
| $\infty$ | $\infty$ | | 25.1327... | | |

TABLE V. For 3-d $O(N)$ $\sigma$ models we present a summary of the analysis of the strong coupling series of $f(\beta)$ (first line) and $f(E)$ (second line) at some values of $N$. We report: the critical point, $\beta_c$ ($E_c$); the estimate of the regular zero of $A_3(\beta)$ ($B_3(E)$) closest to the origin, $\beta_0$ ($E_0$); the values of the approximants $f_{l/m}(\beta)$ ($f_{l/m}(E)$) at $\beta_c$ ($E_c$). Asterisks mark defective PA's. Errors due to the uncertainty of $\beta_c$ and $E_c$ are at most of the order of one in the last digit of the numbers reported (except for some cases where they are given explicitly). We mention that at $N=1$ and $N=2$ our strong coupling analysis led to $E_c = 0.332(3)$ for both.

| $N$ | $\beta_c, E_c$ | $\beta_0, E_0$ | 5/5 | 4/6 | 5/6 | 4/7 | 6/6 | 5/7 | 4/8 |
|---|---|---|---|---|---|---|---|---|---|
| -1 | 0.19840(3) | -0.109 | * | 11.2 | 11.3 | 10.9 | 10.6 | 10.7 | * |
| | 0.350(5) | -0.117 | 9.3 | 9.4 | 9.7 | * | 10.1 | 10.5 | 10.3 |
| 0 | 0.21350(1) [33] | -0.134 | 19.7 | 19.8 | 19.6 | 19.8 | * | 19.4 | * |
| | 0.333(5) | -0.146 | * | 18.1(3) | * | 18.4(2) | 18.3(2) | * | * |
| 1 | 0.221652(4) [34] | -0.149 | 25.4 | 26.0 | 25.3 | 25.8 | 25.4 | 24.8 | * |
| | 0.3301(1) [36] | -0.166 | 24.3 | 24.3 | 24.3 | 24.4 | 24.4 | 24.4 | 24.4 |
| 2 | 0.22710(1) [35] | -0.160 | 29.4 | 29.6 | 29.4 | 29.6 | 29.4 | * | 29.6 |
| | 0.3297(2) [37] | -0.180 | 28.9 | 28.9 | 28.9 | 29.0 | 28.9 | 28.9 | 28.9 |
| 3 | 0.231012(12) [38,39] | -0.168 | 32.5 | 32.3 | * | 32.4 | 32.5 | 32.4 | 32.4 |
| | 0.331(3) | -0.191 | 32.2 | 32.2 | 32.2 | 32.3 | 32.3 | 32.3 | 32.3 |
| 4 | 0.2339(1) | -0.175 | 34.9 | 34.5 | 35.3 | 34.6 | 34.9 | 34.7 | 34.7 |
| | 0.333(2) | -0.200 | 34.8 | 34.8 | 34.7 | 34.9 | 34.8 | 34.9 | 34.8 |
| 8 | 0.2407(1) | -0.19 | 40.5 | 39.9 | 41.1 | 40.2 | 40.6 | 40.4 | 40.3 |
| | 0.334(1) | -0.224 | 40.5 | 40.5 | * | 40.8 | 40.6 | 40.7 | 40.7 |
| 16 | 0.2458(1) | -0.20 | 44.8 | 44.3 | 45.2 | 44.6 | 44.9 | 44.7 | 44.7 |
| | 0.3370(5) | -0.246 | 44.8 | 44.8 | * | 45.1 | 44.9 | 45.0 | 45.0 |
| 24 | 0.2479(1) | -0.21 | 46.5 | 46.1 | 46.8 | 46.4 | 46.6 | 46.4 | 46.4 |
| | 0.3379(3) | -0.260 | 46.4 | 46.5 | 47.0 | 46.7 | 46.6 | 46.7 | 46.6 |
| 32 | 0.2492(2) | -0.22 | 47.5 | 47.1 | 47.6 | 47.3 | 47.5 | 47.4 | 47.4 |
| | 0.3384(3) | -0.268 | 47.3 | 47.4 | 47.7 | 47.5 | 47.5 | 47.5 | 47.5 |
| 48 | 0.2502(1) | -0.22 | 48.4 | 48.1 | 48.5 | 48.3 | 48.4 | 48.3 | 48.3 |
| | 0.3390(3) | -0.280 | 48.2 | 48.3 | 48.5 | 48.5 | 48.4 | 48.4 | 48.4 |
| $\infty$ | 0.252731... | -0.25 | 50.24 | 50.12 | 50.31 | 50.21 | 50.27 | 50.25 | 50.24 |
| | 0.340537... | -0.34 | 50.15 | 50.23 | 50.36 | 50.32 | 50.26 | 50.29 | 50.26 |

TABLE VI. For 3-d $O(N)$ $\sigma$ models we report: $f^*$ as estimated by the analysis of the strong coupling expansion of $f(\beta)$, $f^*_{\rm sc,\beta}$; $f^*$ from the analysis of the series of $f(E)$, $f^*_{\rm sc,E}$; the $O(1/N)$ calculation of $f^*$, $f^*_{1/N}$; $f^*$ from $\phi^4$ field theory at fixed dimensions, $f^*_{\rm ft}$. In Ref. [17] data of $f^*$ were reported without errors, and differences with Refs. [13,14] should be due to a different resummation procedure.

| $N$ | $f^*_{\rm sc,\beta}$ | $f^*_{\rm sc,E}$ | $f^*_{1/N}$ | $f^*_{\rm ft}$ |
|---|---|---|---|---|
| -1 | 10.7(4) | 10.3(6) | | |
| 0 | 19.4(3) | 18.3(3) | | 17.86(5) [14,13] 17.62 [17] |
| 1 | 25.1(5) | 24.4(1) | | 23.72(8) [14,13] 23.47 [17] |
| 2 | 29.5(1) | 28.9(1) | | 28.27(8) [14,13] 28.03 [17] |
| 3 | 32.4(1) | 32.3(1) | | 31.78(9) [14,13] 31.60 [17] |
| 4 | 34.8(3) | 34.8(1) | 30.84 | 34.41 [17] |
| 8 | 40.4(4) | 40.7(1) | 40.55 | 40.93 [17] |
| 16 | 44.8(3) | 45.0(1) | 45.41 | 45.50 [17] |
| 24 | 46.5(2) | 46.6(2) | 47.03 | 47.13 [17] |
| 32 | 47.4(2) | 47.5(1) | 47.84 | 47.94 [17] |
| 48 | 48.3(1) | 48.4(1) | 48.65 | |
| $\infty$ | 50.25(6) | 50.27(6) | 50.2654... | |

TABLE VII. We report the values of $\beta_n$, defined in Eq. (A5), in $d = 1, 2, 3$ such that $\beta_n \gtrsim 10^{-3}$. Notice that in 2-d: $\beta_1 = \frac{44}{3} + \frac{128}{27}\pi^2 - \frac{64}{9}\psi'(1/3) = -10.33501055$.

| | $d = 1$ | $d = 2$ | $d = 3$ |
|---|---|---|---|
| $\beta_1$ | $-\frac{388}{27}$ | $-10.33501055$ | $-\frac{164}{27}$ |
| $\beta_2$ | $\frac{1187}{108}$ | 5.00027593 | 1.34894276 |
| $\beta_3$ | $-\frac{335}{162}$ | $-0.08884297$ | 0.15564589 |
| $\beta_4$ | $-\frac{10001}{46656}$ | $-0.00407962$ | 0.05123618 |
| $\beta_5$ | $-\frac{605}{11664}$ | 0.00506747 | 0.02342417 |
| $\beta_6$ | $-\frac{20045}{1119744}$ | 0.00491122 | 0.01264064 |
| $\beta_7$ | $-\frac{38671}{5038848}$ | 0.00377364 | 0.00757889 |
| $\beta_8$ | $-\frac{1231807}{322486272}$ | 0.00281096 | 0.00489401 |
| $\beta_9$ | $-\frac{21367}{10077696}$ | 0.00211235 | 0.00334024 |
| $\beta_{10}$ | $-\frac{89062753}{69657034752}$ | 0.00161697 | 0.00237987 |
| $\beta_{11}$ | $-\frac{28651973}{34828517376}$ | 0.00126267 | 0.00175481 |
| $\beta_{12}$ | | 0.00100476 | 0.00133070 |
| $\beta_{13}$ | | 0.00081329 | 0.00103290 |
| $\beta_{14}$ | | | 0.00081770 |

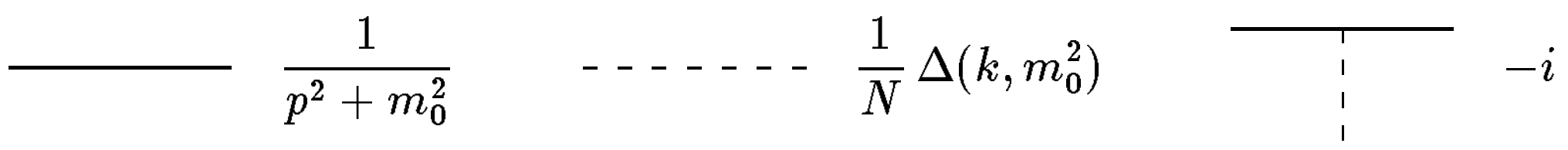

FIG. 1.

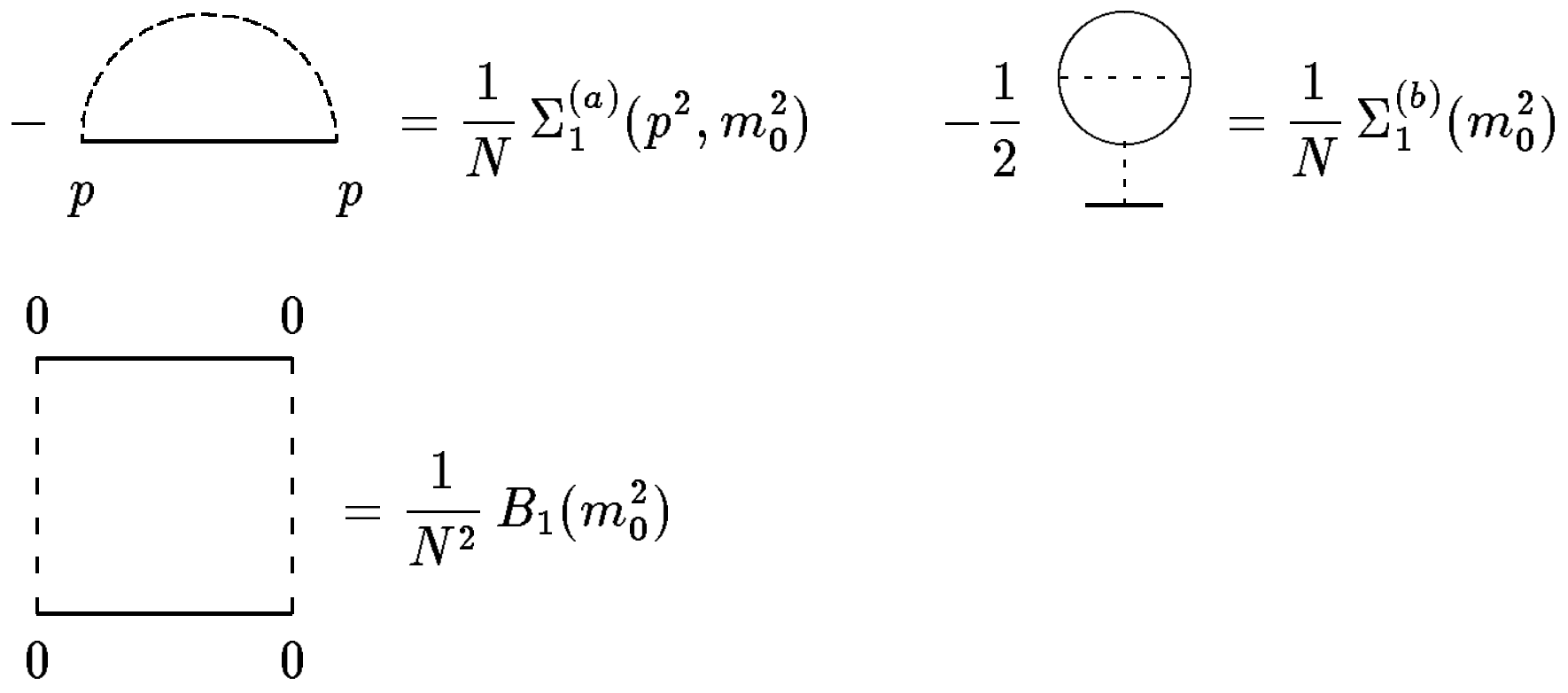

FIG. 2.

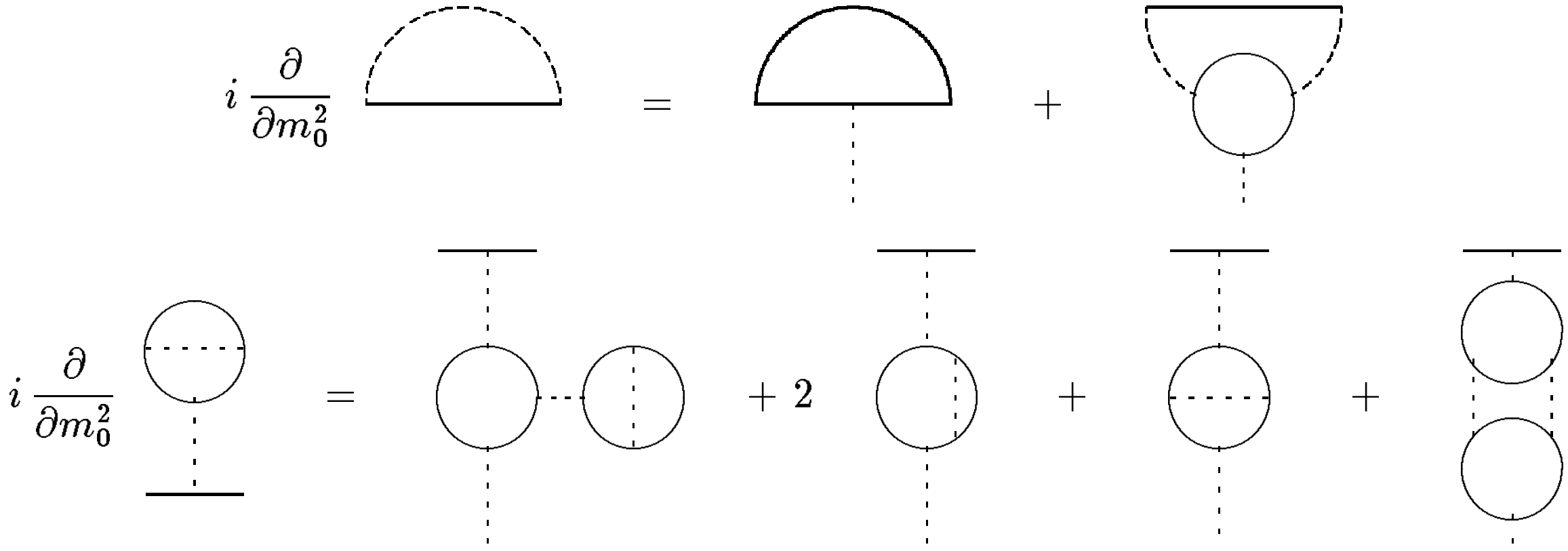

FIG. 3.

Figure 4

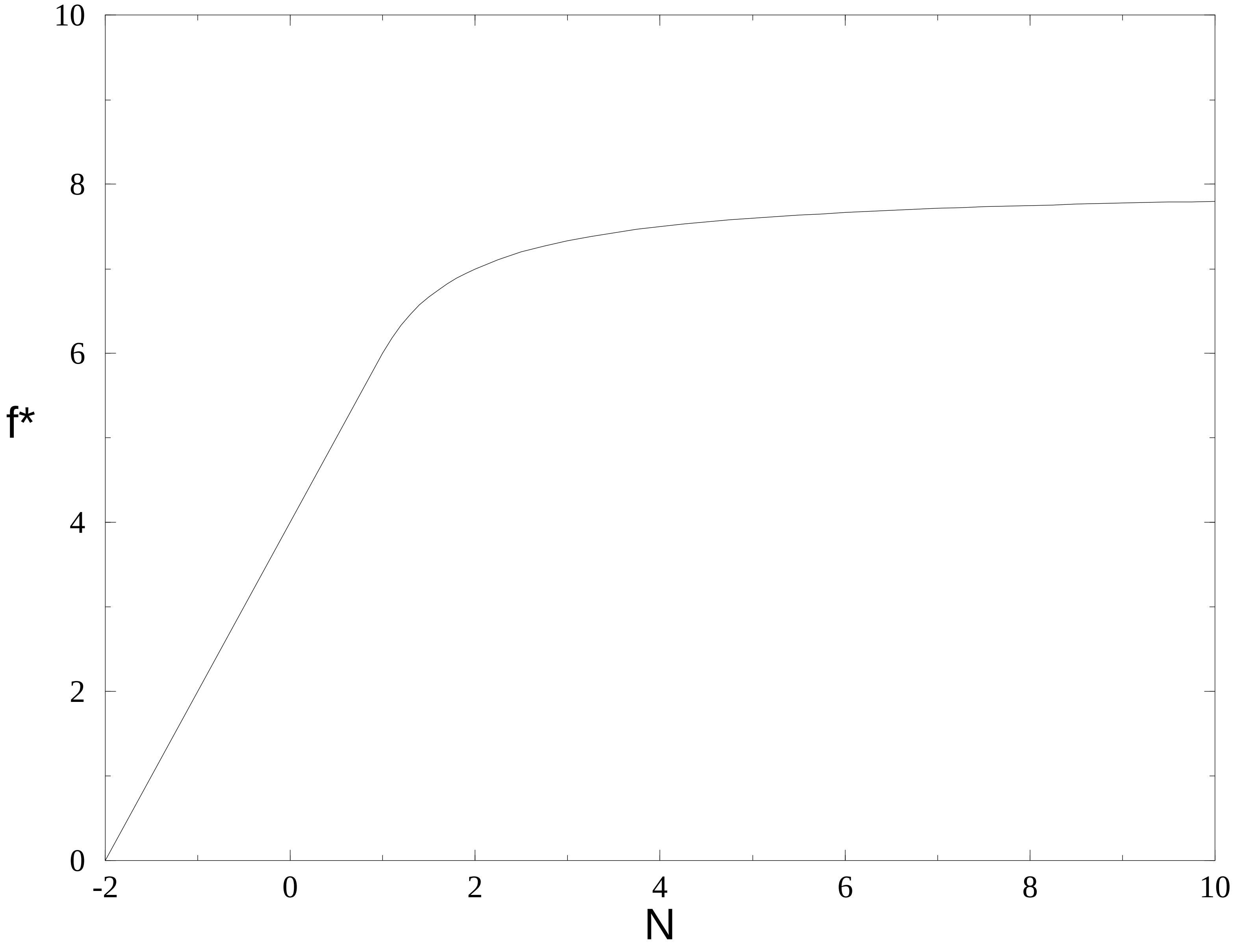

# Figure 5

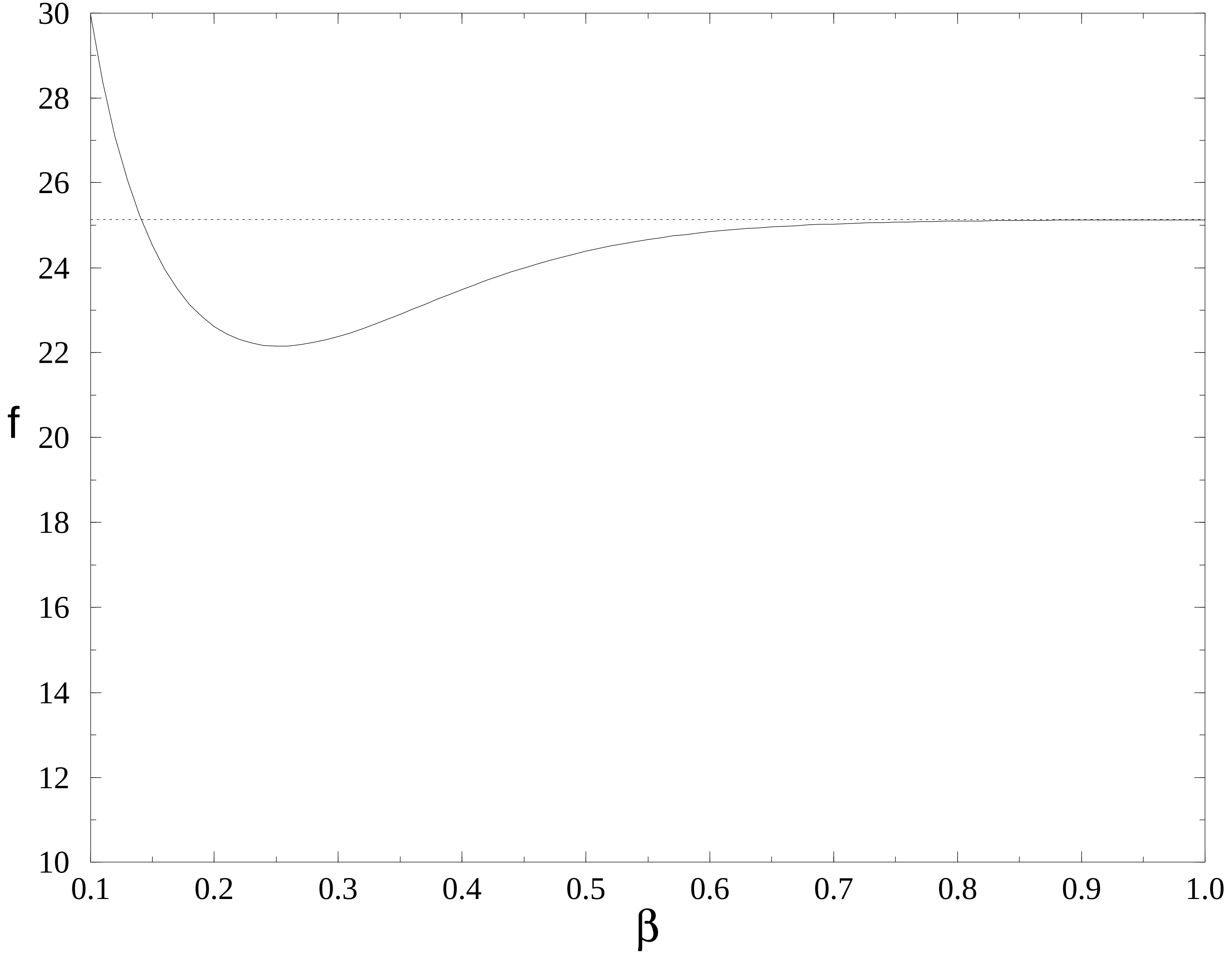

# Figure 6

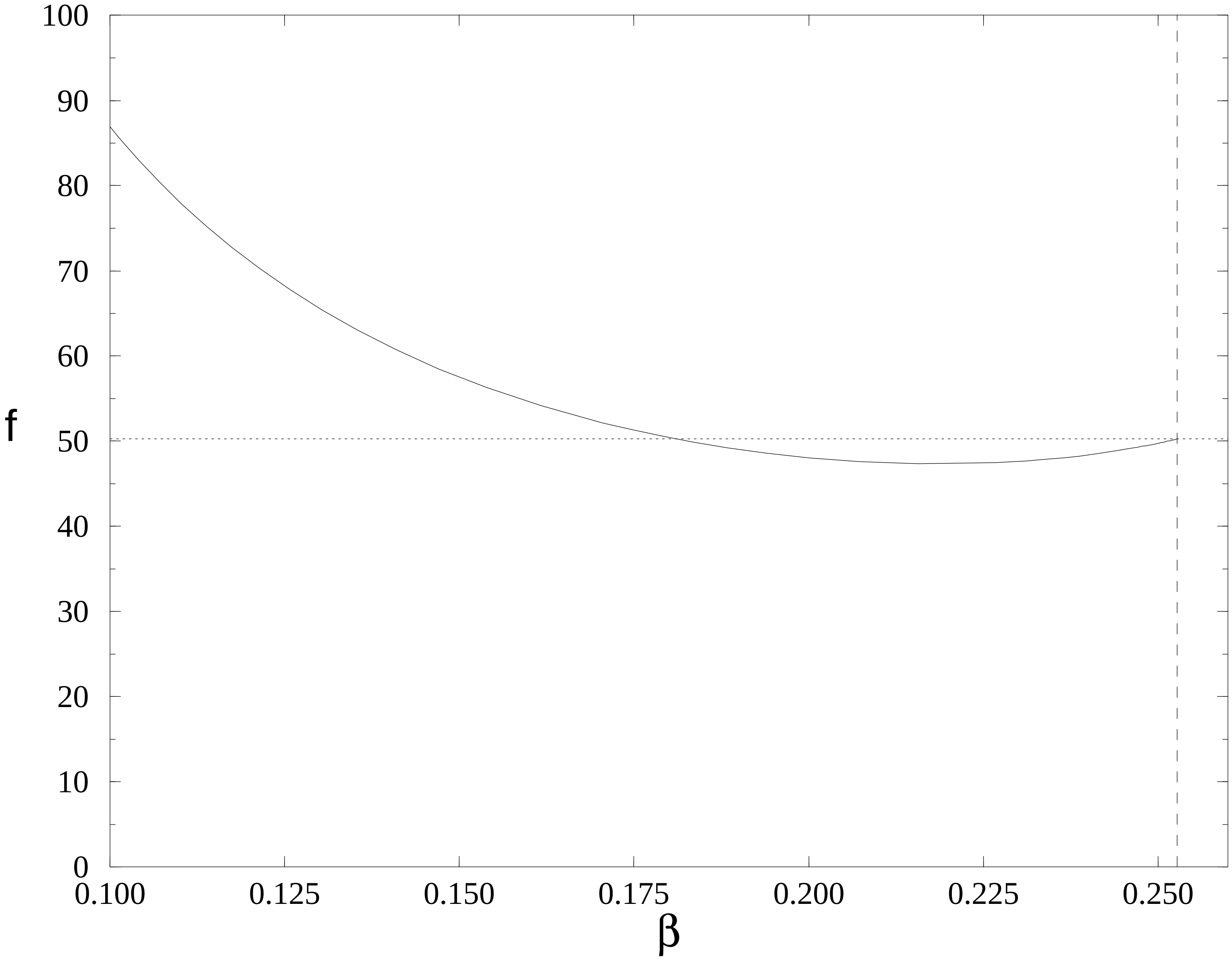

# Figure 7

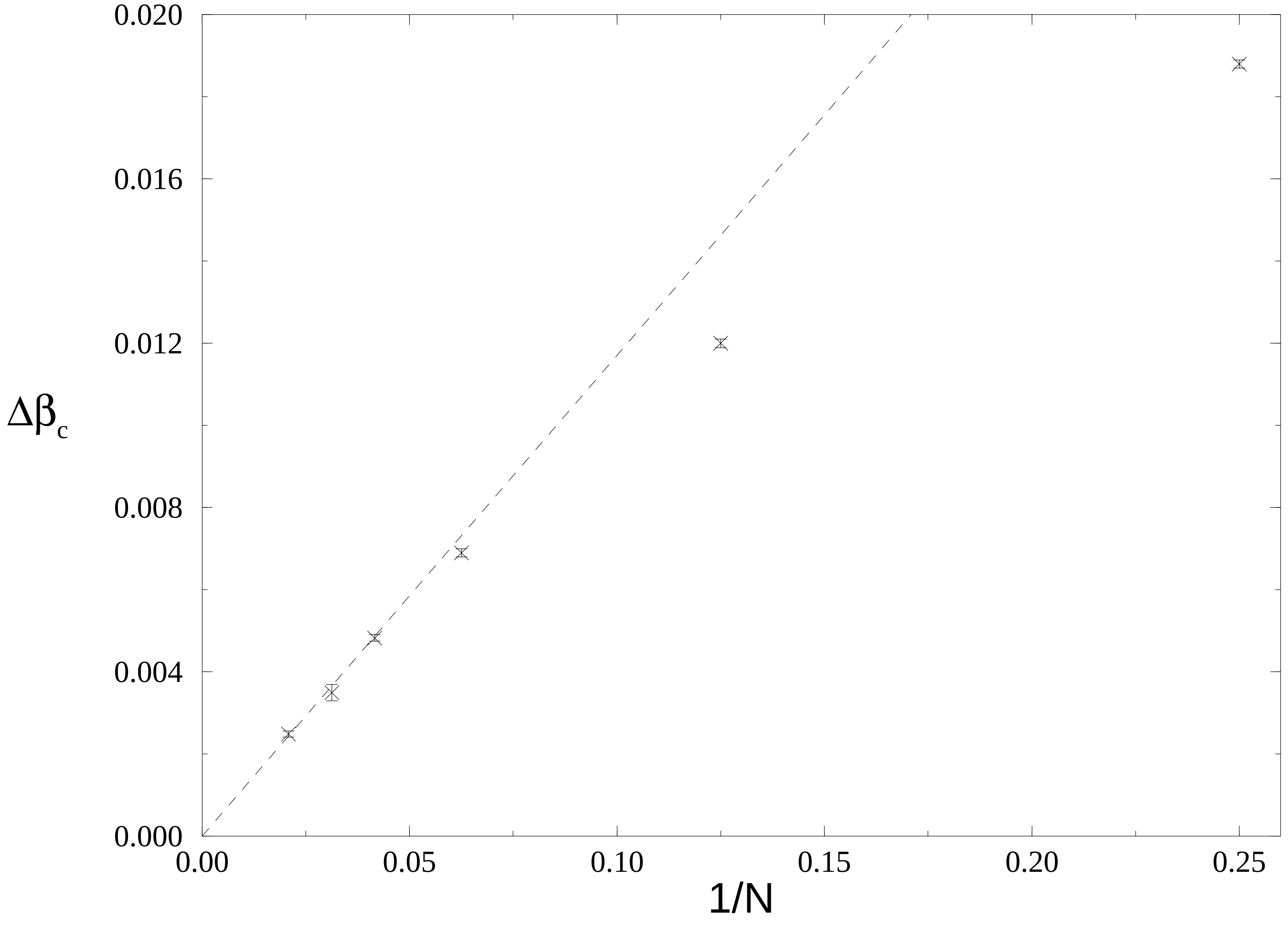

Figure 8

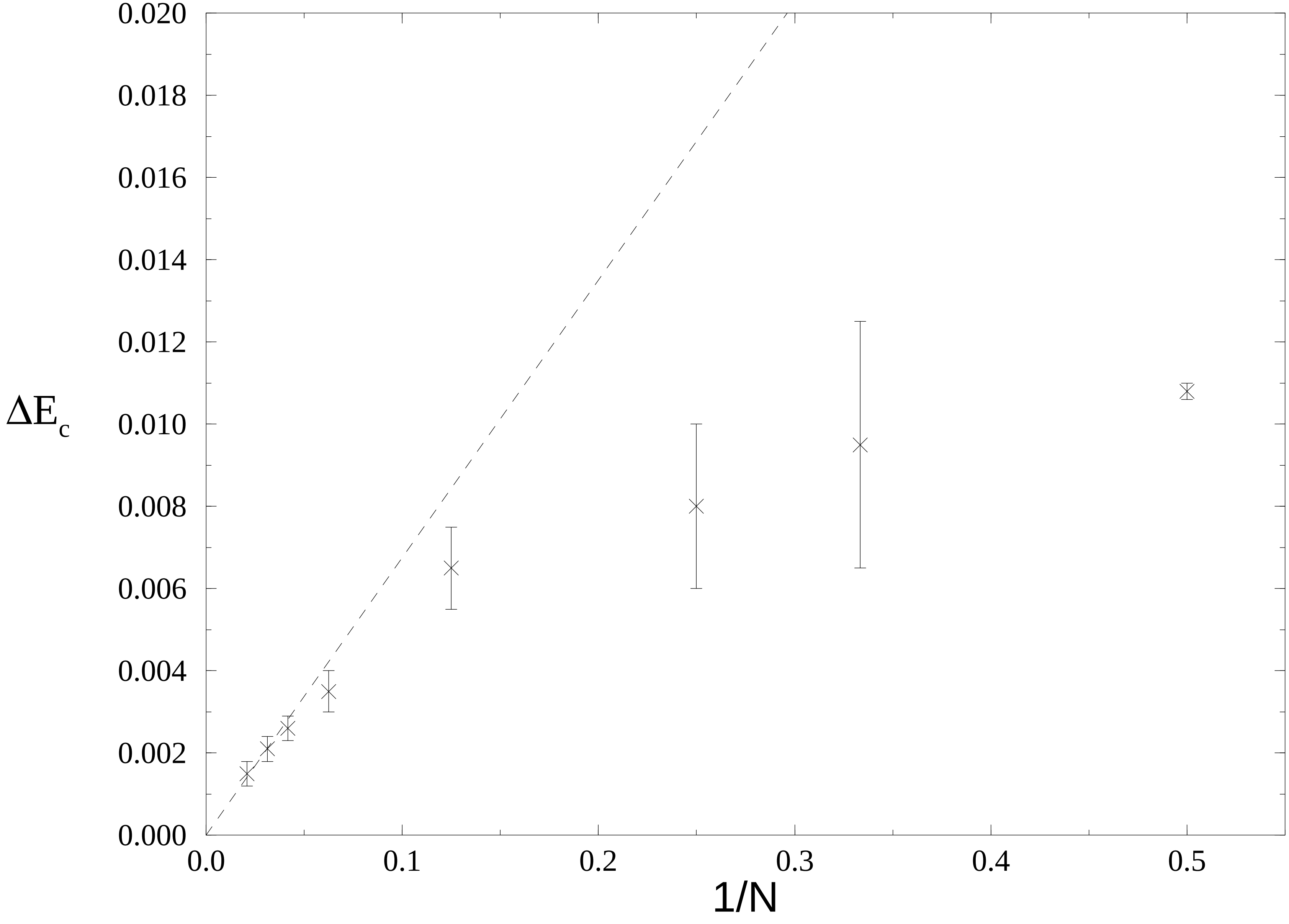

Figure 9

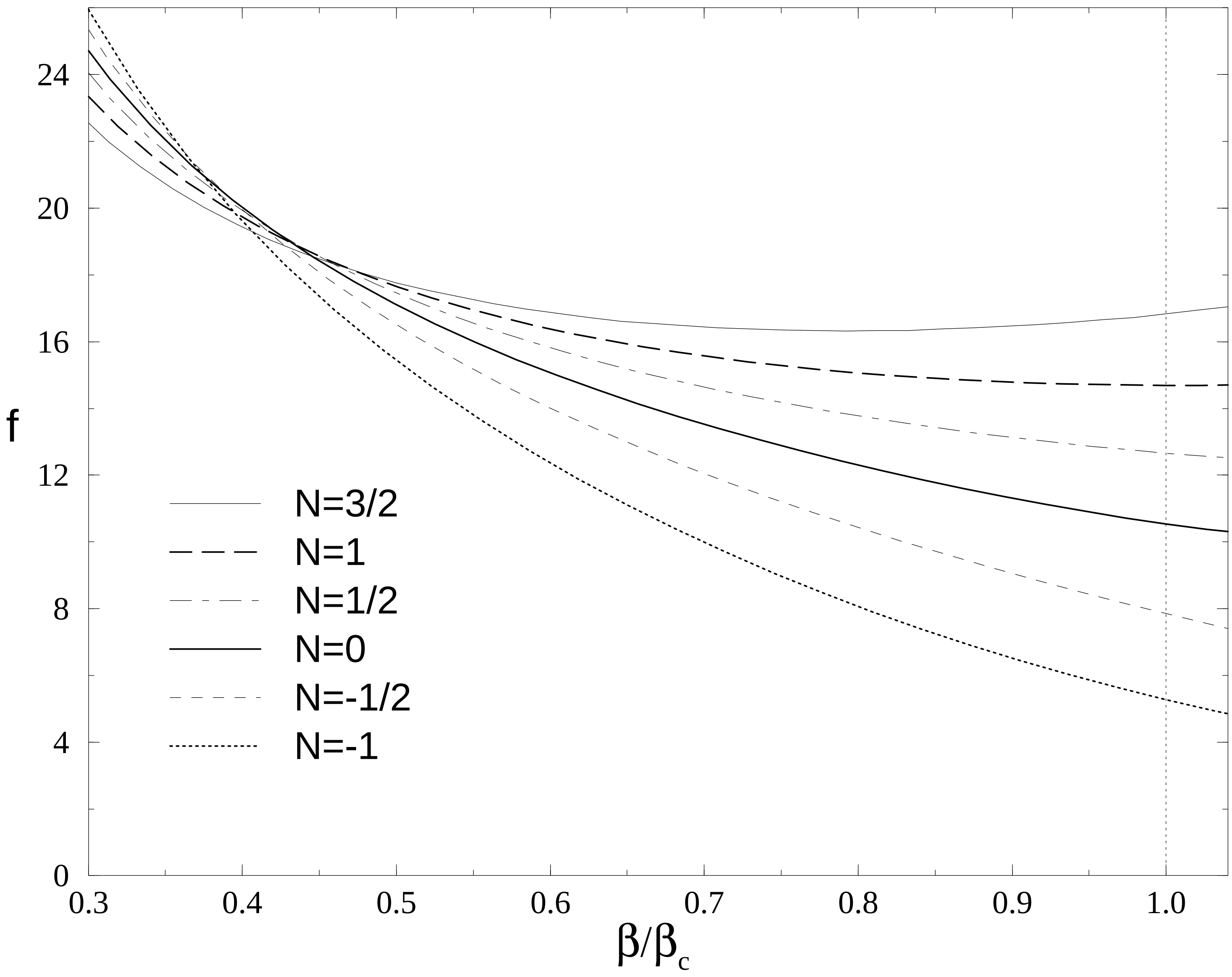

# Figure 10

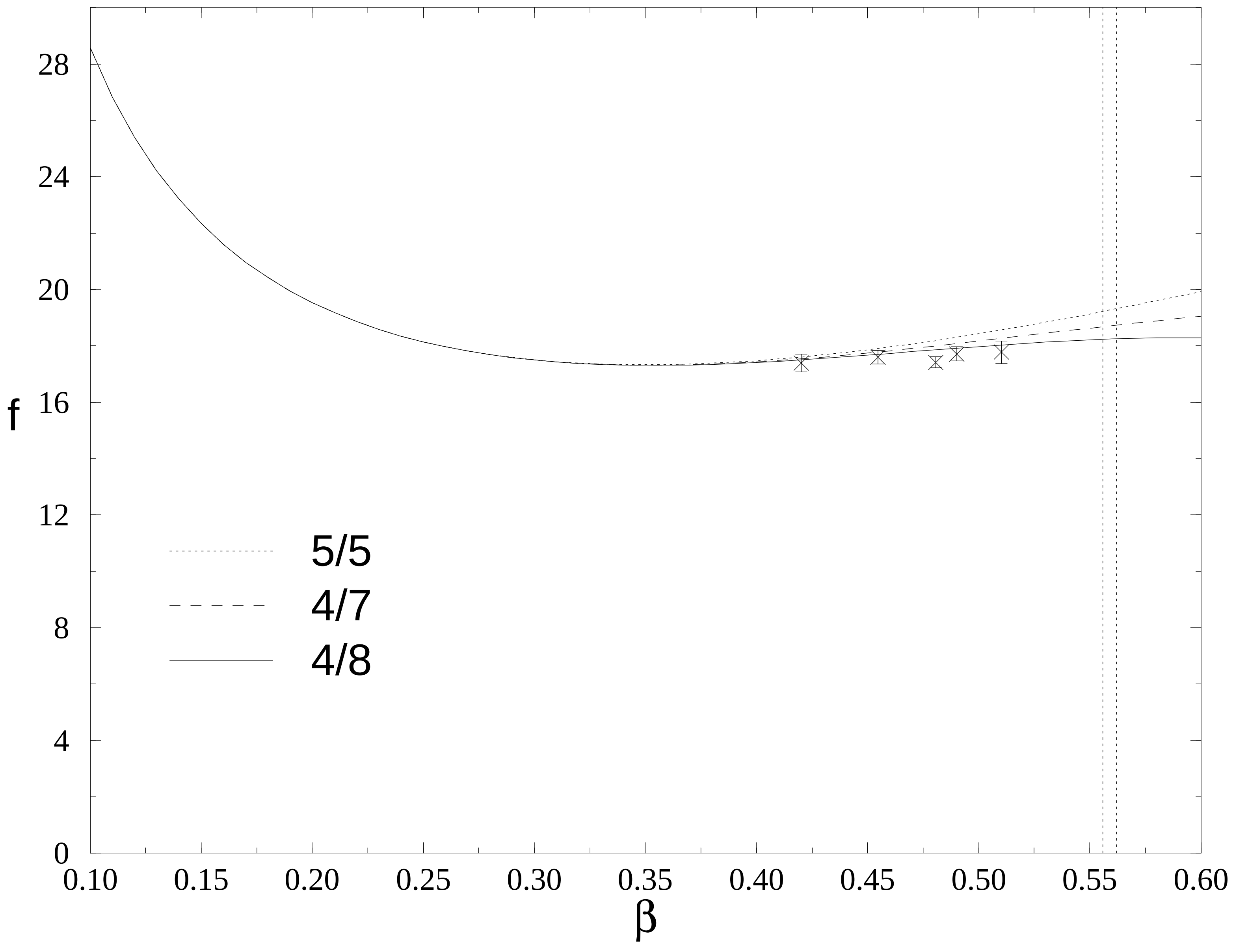

# Figure 11

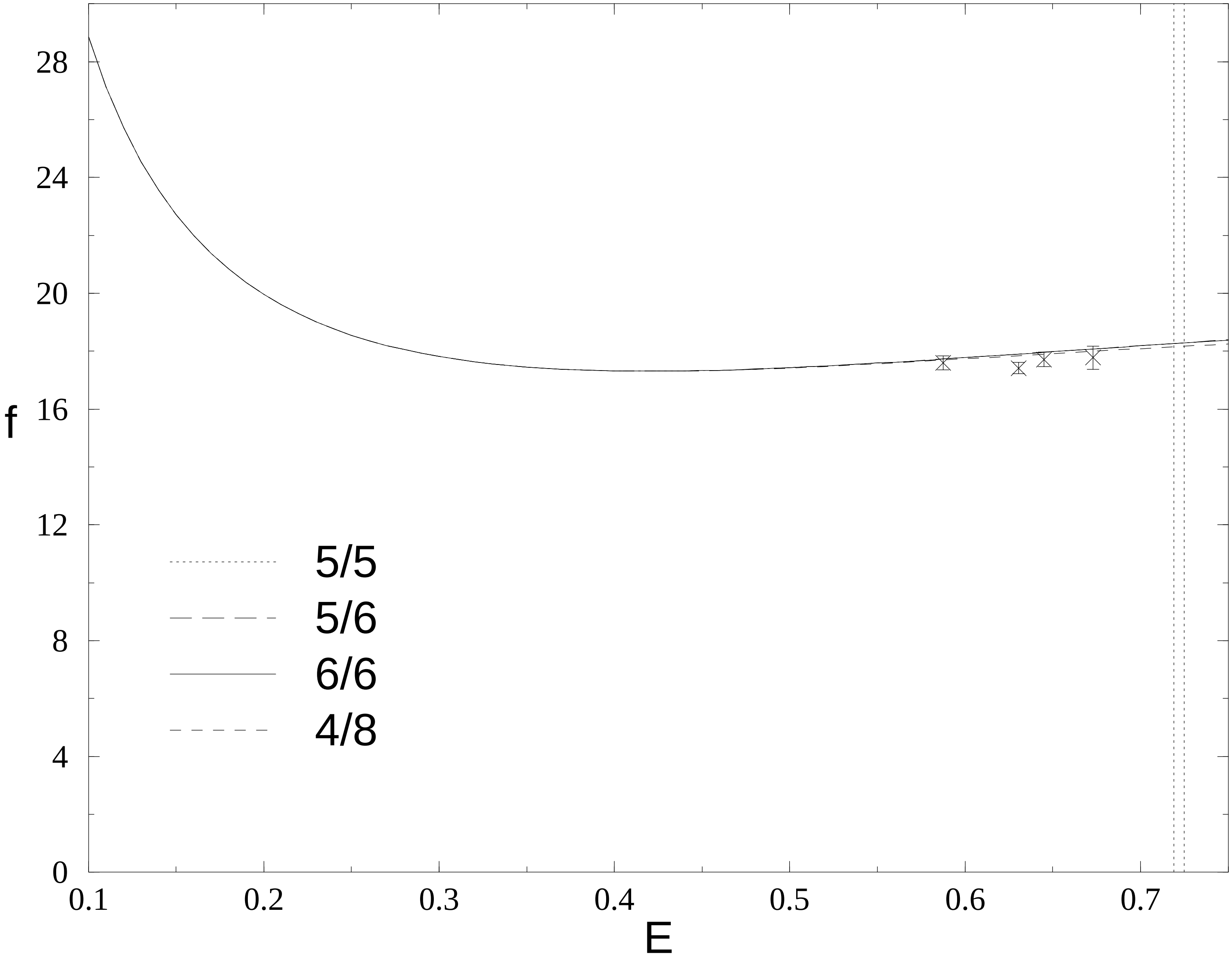

Figure 12

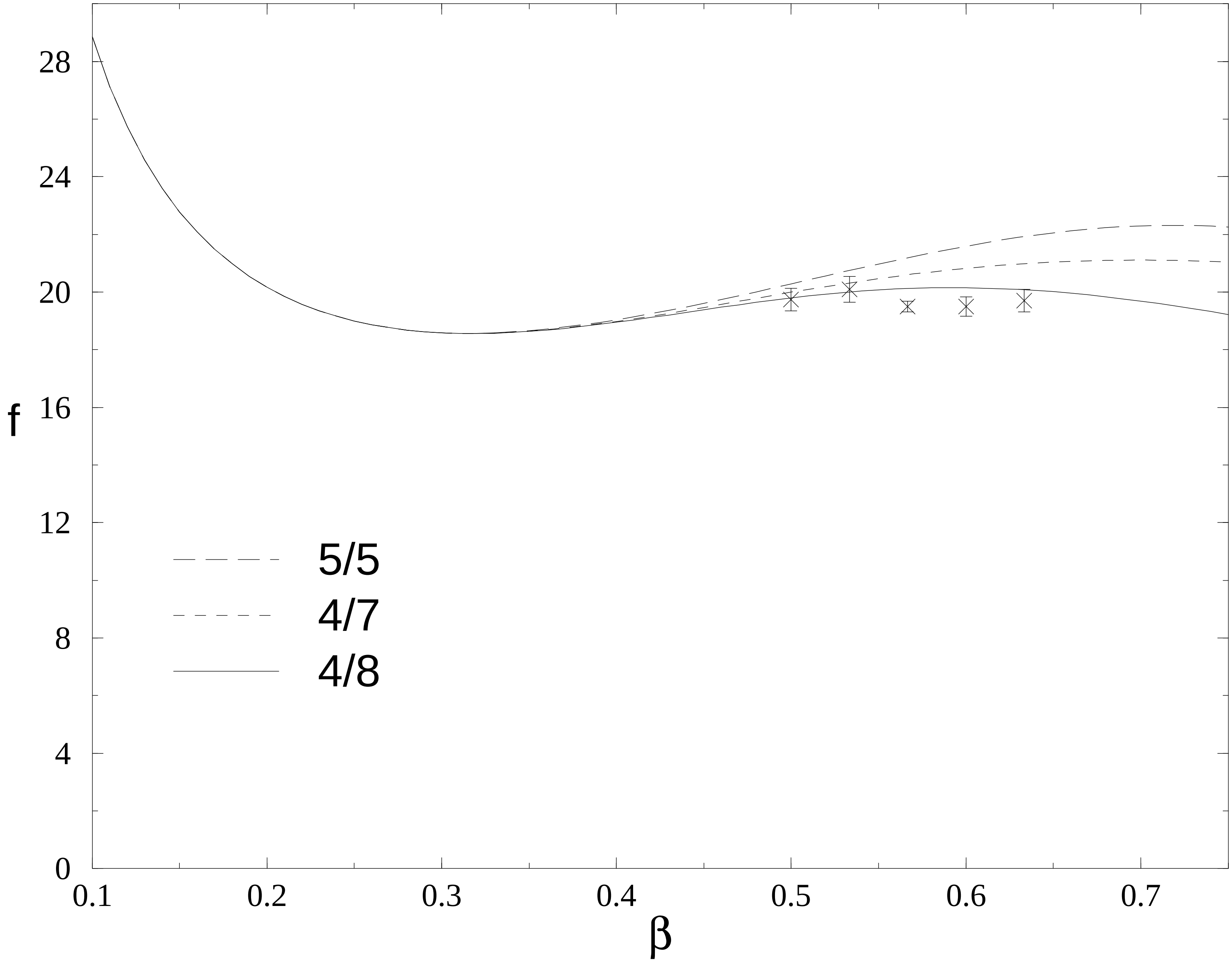

# Figure 13

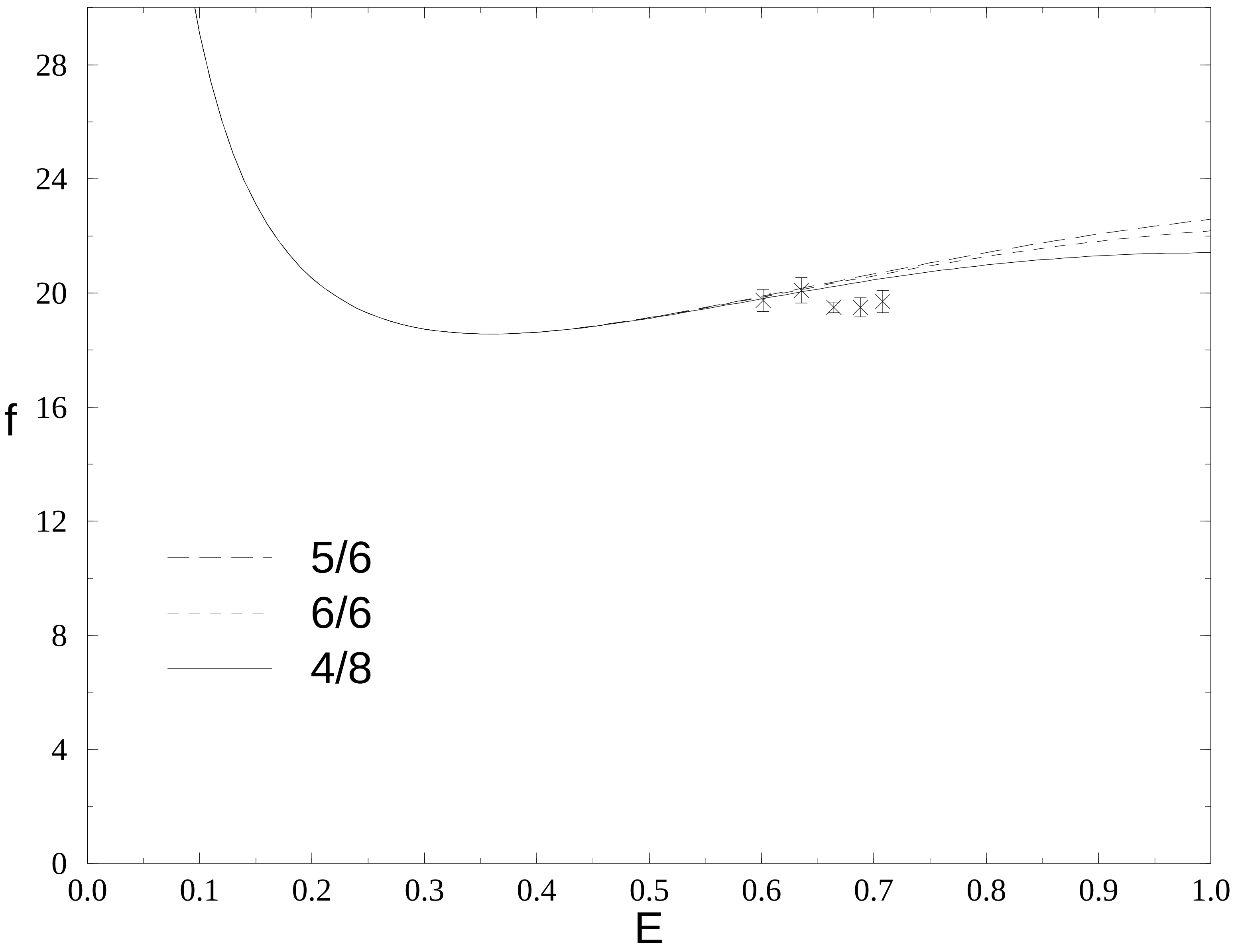

Figure 14

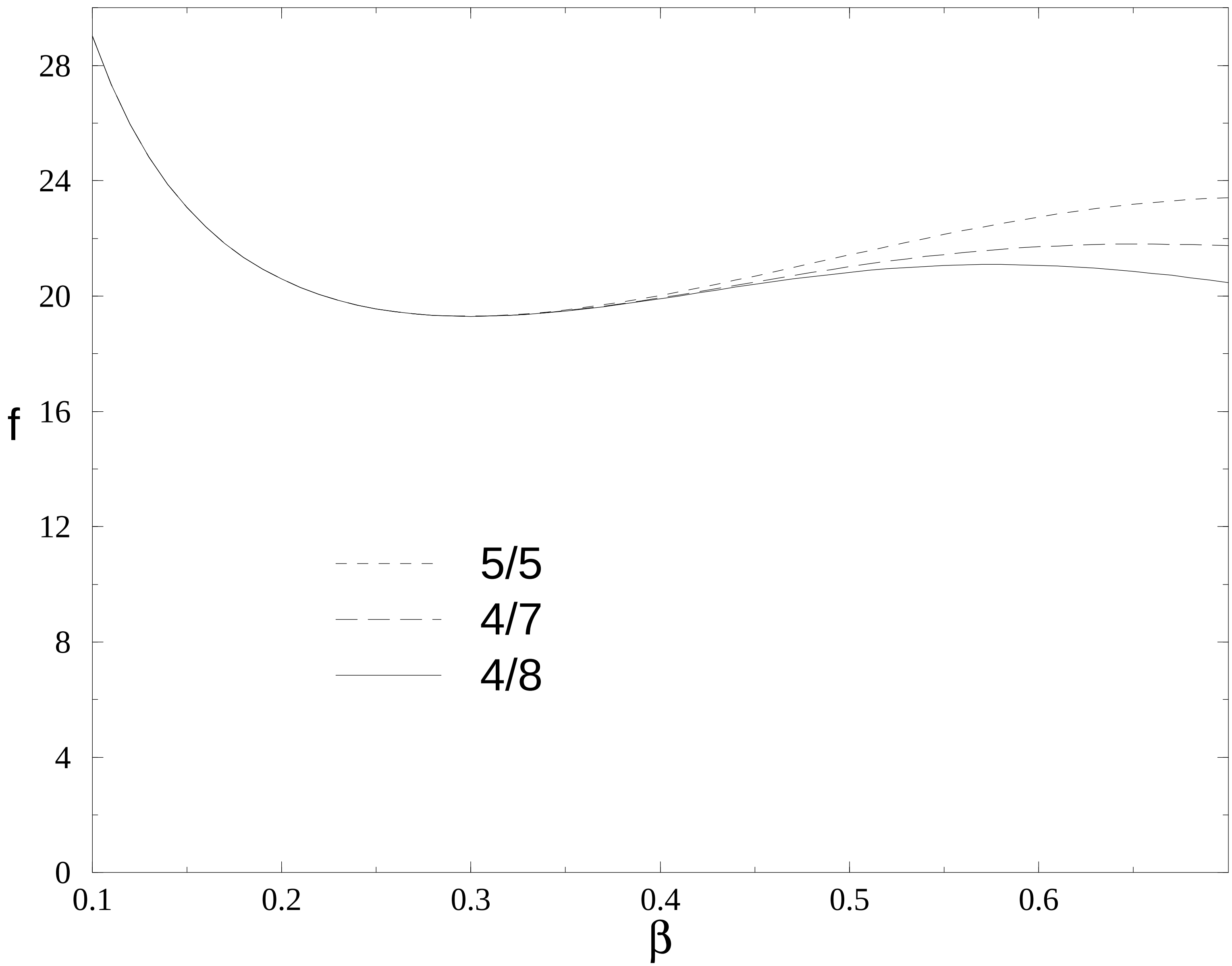

Figure 15

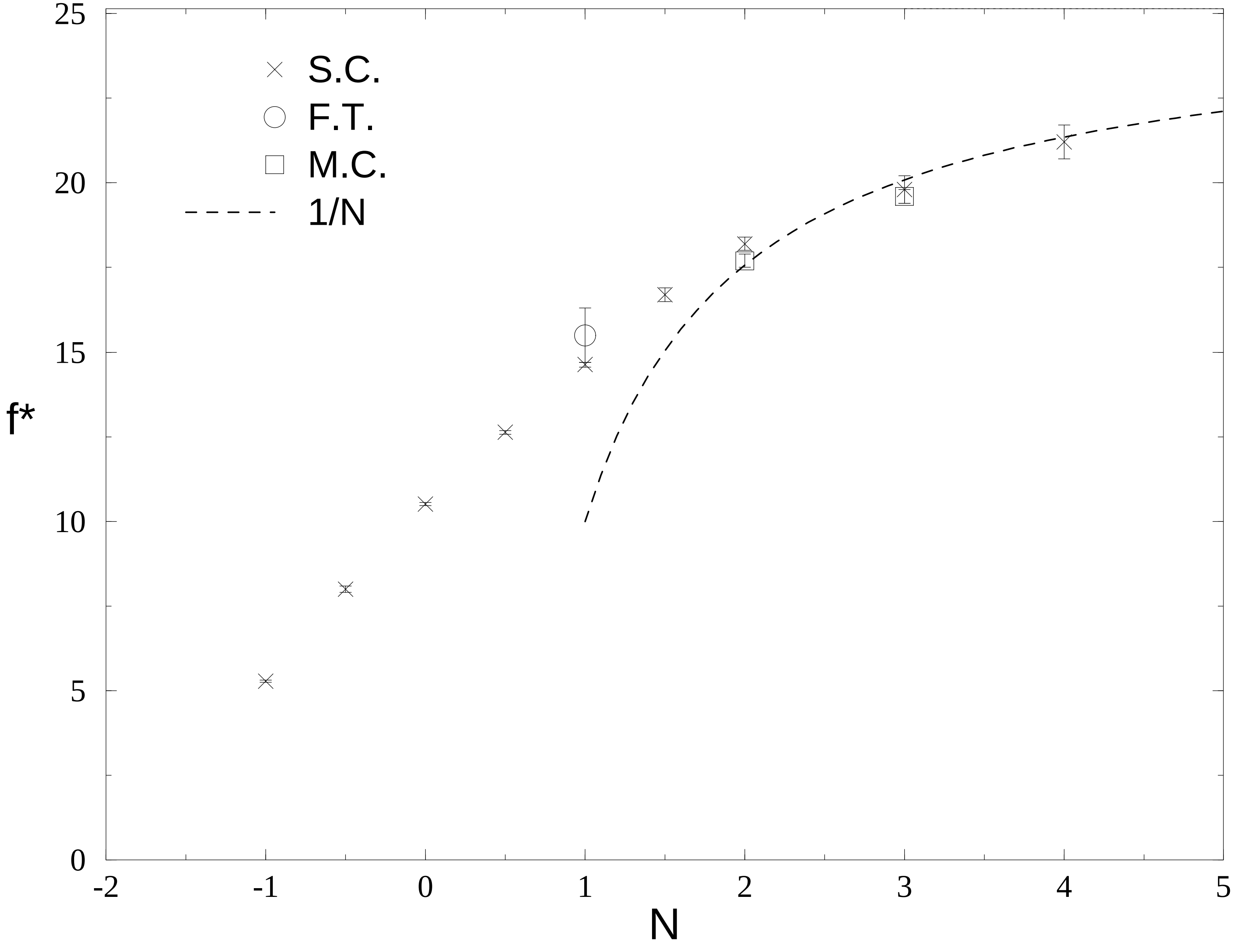

Figure 16

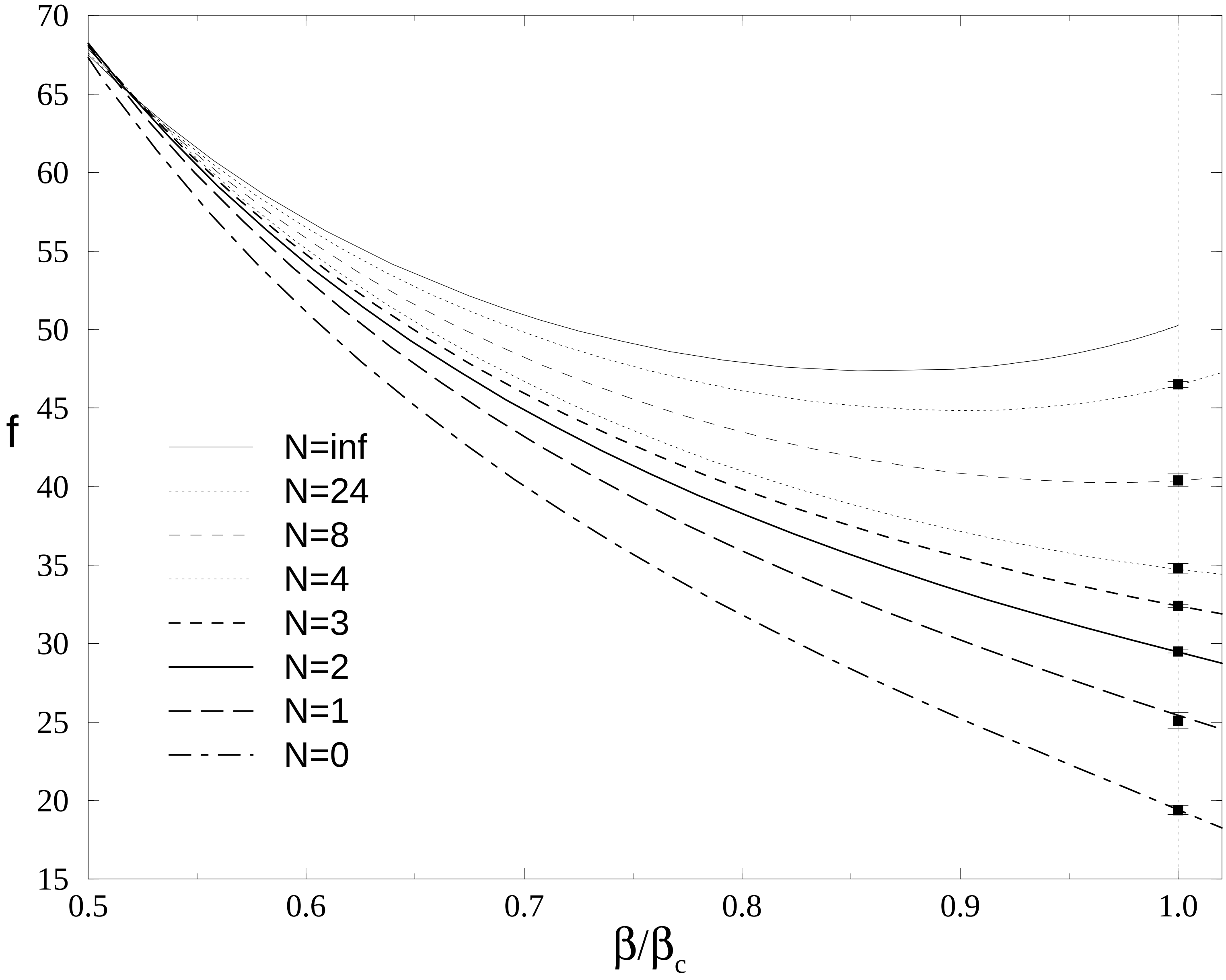

Figure 17

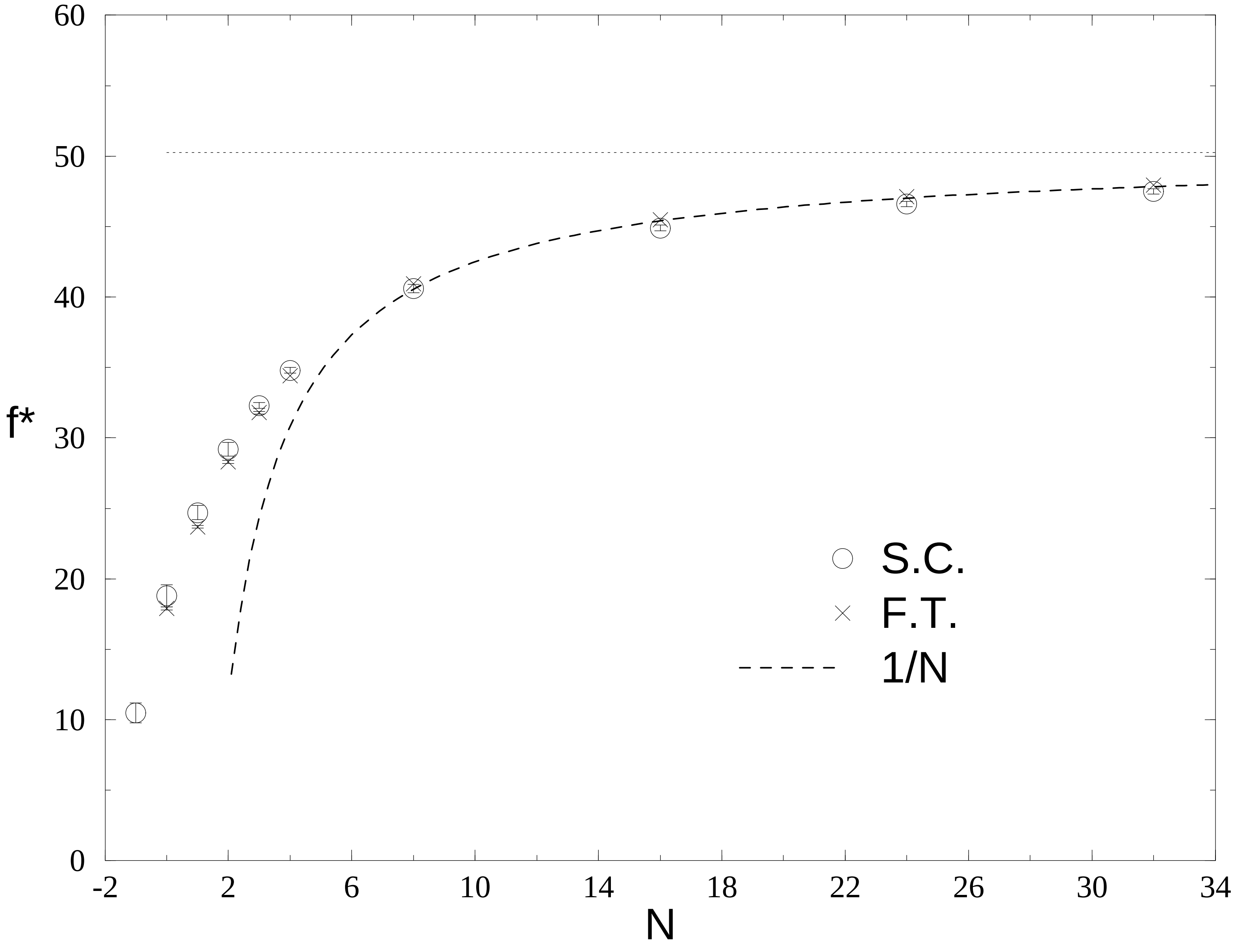

Figure 18

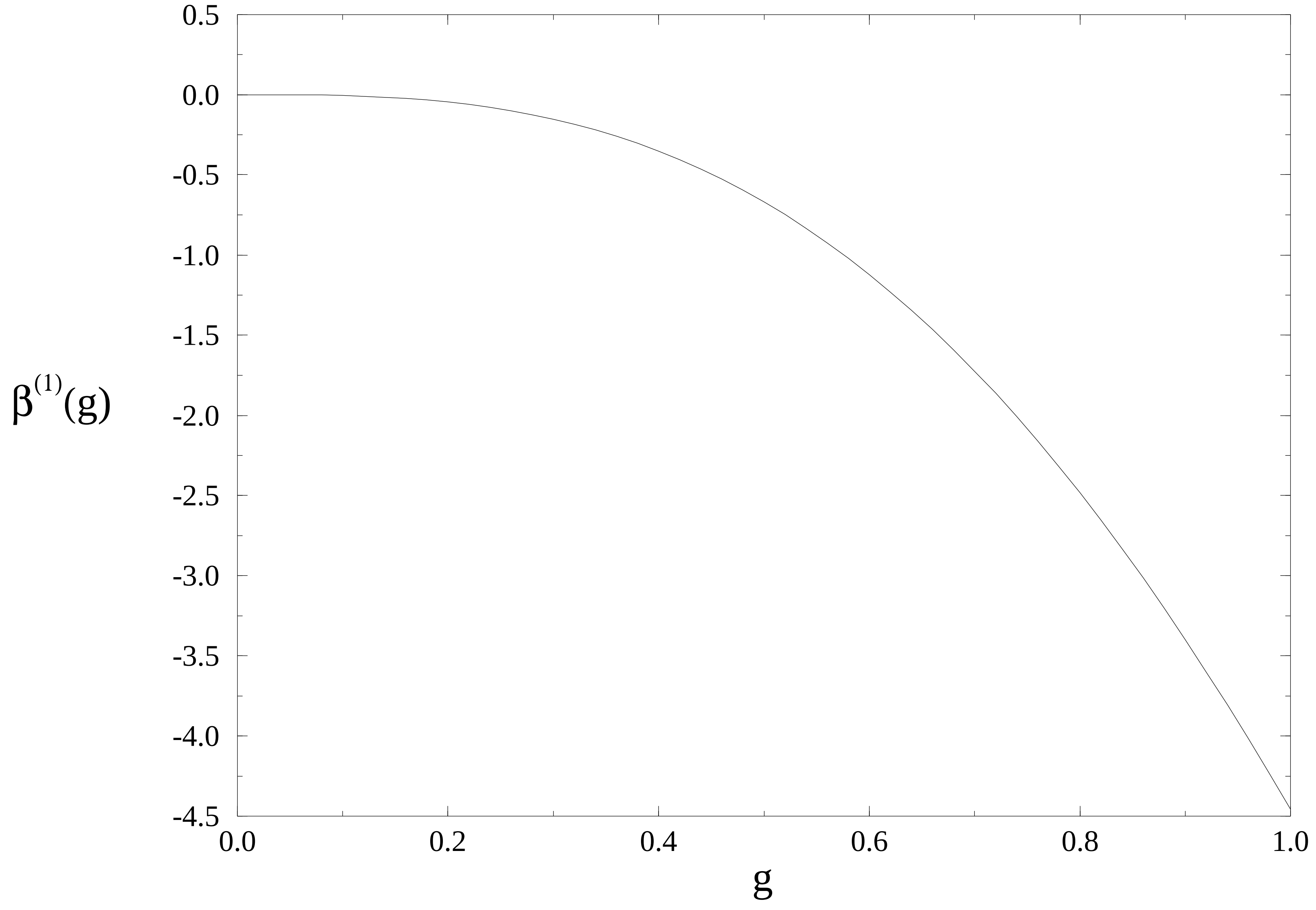

Figure 19

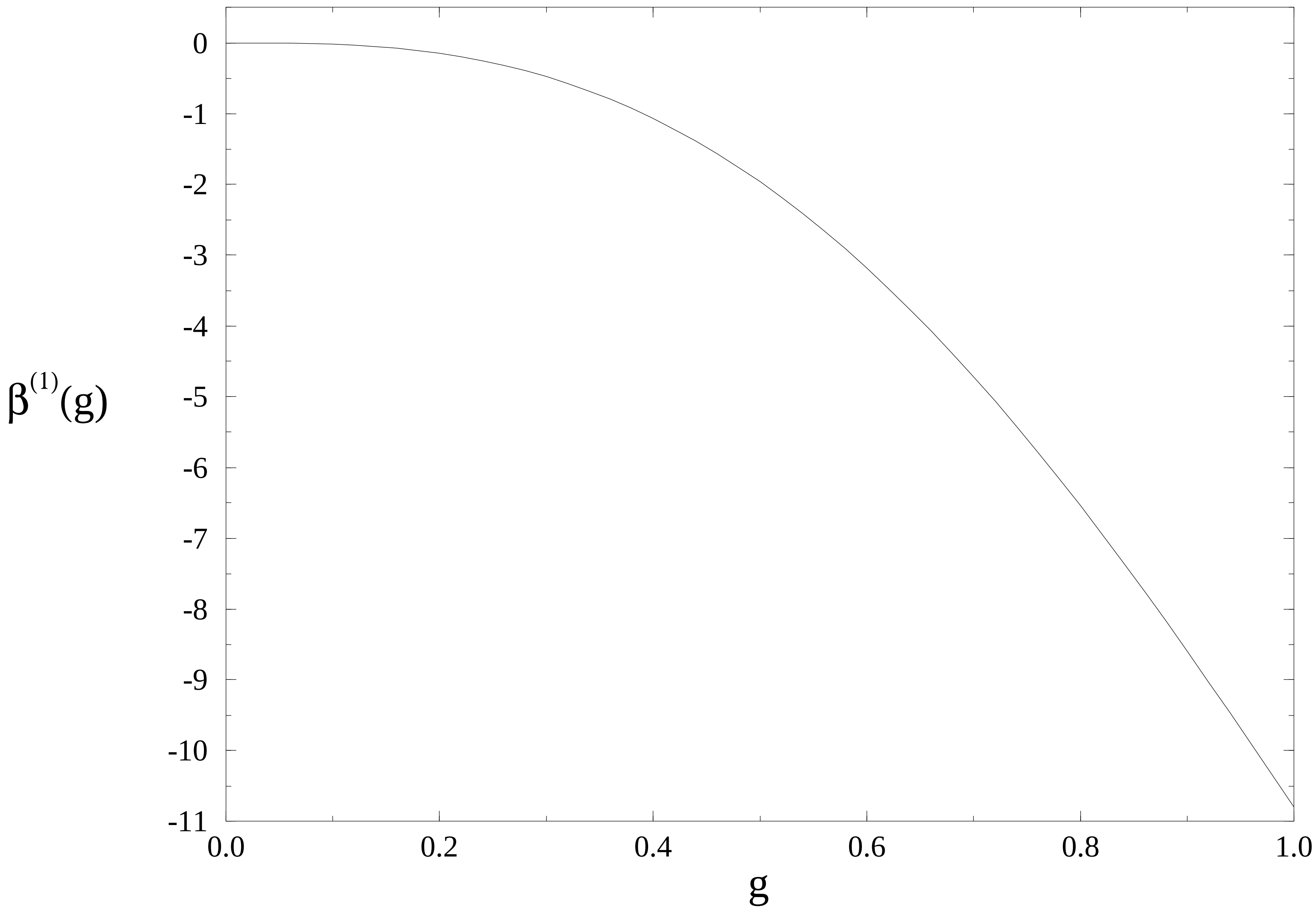